\documentclass[12pt]{article}
\usepackage{array}
\usepackage{graphicx,epsfig}
\usepackage{amsmath,amssymb,amsthm,bm,bbm}
\usepackage{verbatim}
\usepackage{subfigure}
\usepackage[english]{babel}
\usepackage{latexsym,psfrag,array,multicol,palatino,enumerate,caption,multirow}
\usepackage{setspace}
\usepackage[usenames]{color}
\usepackage{dsfont}
\usepackage{multirow}
\usepackage{hyperref}
\usepackage[numbers,sort&compress]{natbib}

\newcommand{\blind}{0}

\date{}
\begin{document}
\def\spacingset#1{\renewcommand{\baselinestretch}%
{#1}\small\normalsize} \spacingset{1}


\if0\blind
{
  \title{\bf A Bayesian State-Space Approach to Mapping Directional Brain Networks}

\author{Huazhang Li\thanks{Co-first authors.}\\
Department of Statistics, University of Virginia\\
Yaotian Wang\footnotemark[1]\\
Department of Statistics, University of Pittsburgh\\
Guofen Yan\\
Department of Public Health Sciences, University of Virginia\\
Yinge Sun\\
Department of Statistics, University of Virginia\\
Seiji Tanabe\\
Department of Psychology, University of Virginia\\
Chang-Chia Liu\\
Department of Neurological Surgery, University of Virginia\\
Mark Quigg\\
Department of Neurology, University of Virginia\\
Tingting Zhang\thanks{The corresponding author e-mail address: TIZ67@pitt.edu}\thanks{The author gratefully acknowledge the support by NSF-1758095.}\\
Department of Statistics, University of Pittsburgh}
 \maketitle
} \fi

\if1\blind
{
  \bigskip
  \bigskip
  \bigskip
  \begin{center}
    {\LARGE\bf A Bayesian State-Space Approach to Mapping Directional Brain Networks}
\end{center}
  \medskip
} \fi

\bigskip
\begin{abstract}
The human brain is a directional network system of brain regions involving directional connectivity. Seizures are a directional network phenomenon as abnormal neuronal activities start from a seizure onset zone (SOZ) and propagate to otherwise healthy regions. To localize the SOZ of an epileptic patient, clinicians use iEEG to record the patient's intracranial brain activity in many small regions. iEEG data are high-dimensional multivariate time series. We build a state-space multivariate autoregression (SSMAR) for iEEG data to model the underlying directional brain network. To produce scientifically interpretable network results, we incorporate into the SSMAR the scientific knowledge that the underlying brain network tends to have a cluster structure. Specifically, we assign to the SSMAR parameters a stochastic-blockmodel-motivated prior, which reflects the cluster structure. We develop a Bayesian framework to estimate the SSMAR, infer directional connections, and identify clusters for the unobserved network edges. The new method is robust to violations of model assumptions and outperforms existing network methods. By applying the new method to an epileptic patient's iEEG data, we reveal seizure initiation and propagation in the patient's brain network. Our method can also accurately localize the SOZ. Overall, this paper provides a tool to study the human brain network.
\end{abstract}


\vfill

\newpage
\spacingset{1.5} 

\section{Introduction}
Brain activities form a directional network, where network nodes are brain regions and each network edge represents a directional influence exerted by one region on another. Such directional information flow from one region to another is referred to as directional connectivity also called effective connectivity \citep{Friston94}. The purposes of this paper are to present a new statistical approach for analysis of intracranial electroencephalographic (iEEG) data and to use our approach to uncover the normal and abnormal directional brain networks of epileptic patients over the course of seizure development.

\begin{figure}[h]
\centering
\subfigure[iEEG electrode grid]
{\label{fig:Sub1}
    \includegraphics[width=5.1cm,height=3.8cm,trim= 60mm 20mm 85mm 15mm,clip=TRUE]{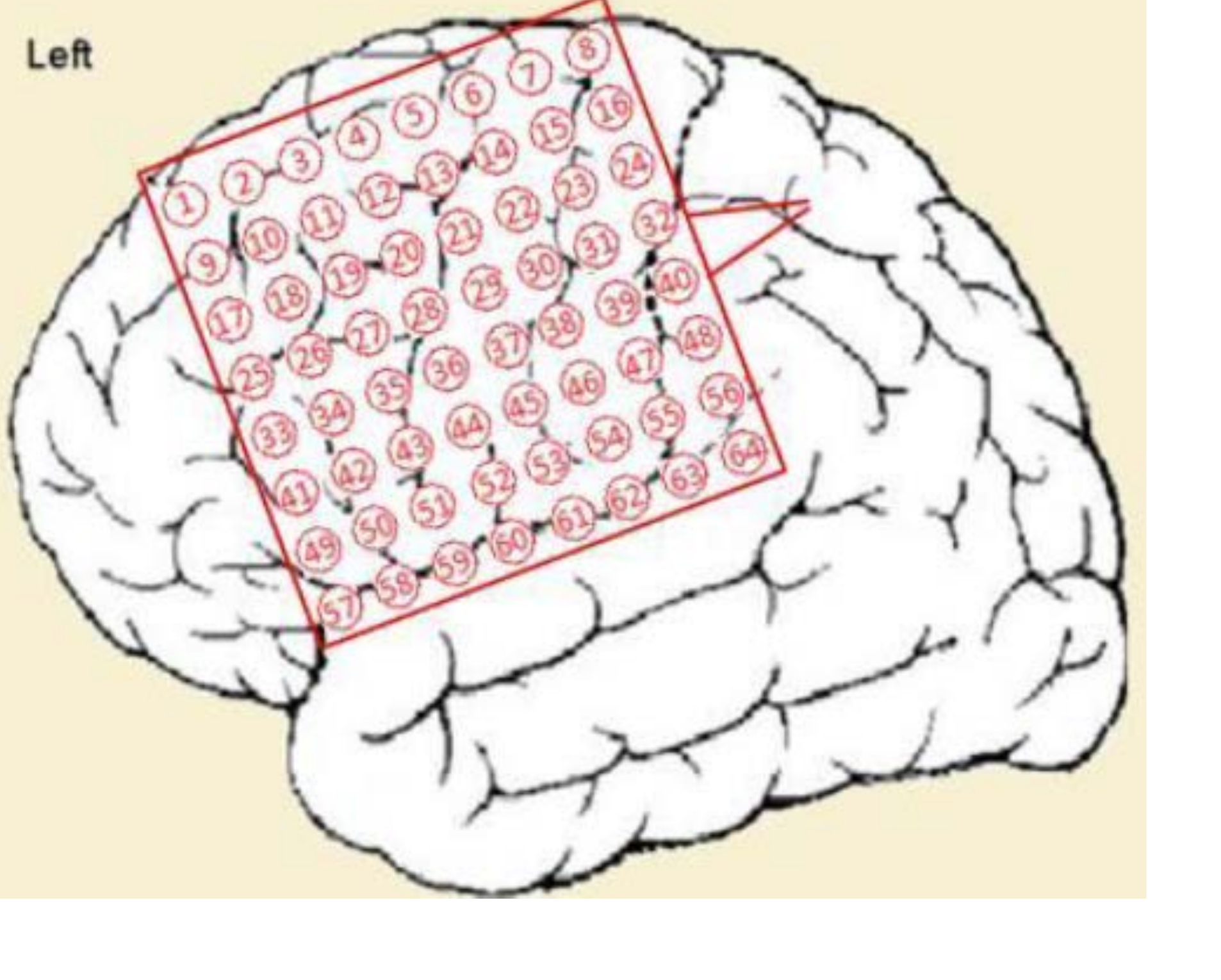}
    }
   \subfigure[5-second iEEG recordings from 2 channels]
   {\label{fig:ECoGSeries100}
    \includegraphics[width=7.1cm,height=4.0cm,trim= 18mm 0mm 18mm 5mm,clip=TRUE]{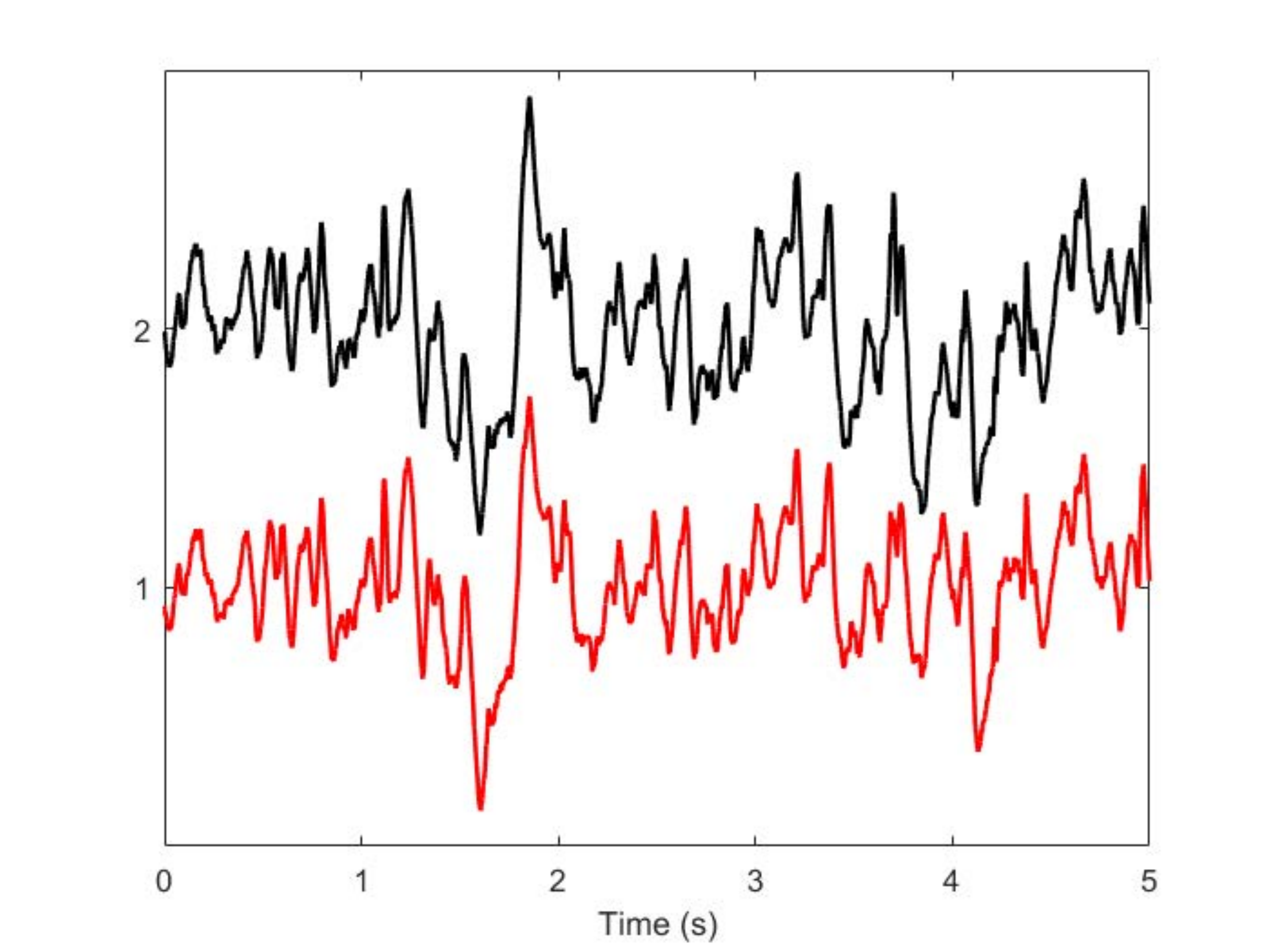}
    }
    \caption{(a) The iEEG electrode placement on the left hemisphere of the epileptic patient under study. (b) iEEG time-series segments of two regions/channels.}
\end{figure}

Seizures are a directional network phenomenon \citep{rosenow2001presurgical}, as abnormal, excessive, and synchronous neuronal activities start from the seizure onset zone (SOZ) and propagate to otherwise healthy brain regions. Brain surgery to remove the SOZ is a common treatment consideration for patients with drug resistant epilepsy. Pre-surgical evaluation includes localization of the SOZ using iEEG, which is absolutely critical to the success of the surgery. Clinicians place iEEG electrodes on the exposed brain (inside the skull) of epileptic patients to record their neuronal activities in many regions. The recorded data are high-dimensional multivariate time-series of voltage waveforms, which often exceed 50 channels (with each channel corresponding to one region). Figure \ref{fig:Sub1} shows the electrode placement on the left hemisphere of a patient who underwent iEEG recordings in epilepsy evaluation. Figure \ref{fig:ECoGSeries100} illustrates 5-second segments of the patient's iEEG recordings in two regions/channels.

To localize the SOZ, trained EEG experts visually examine iEEG waveforms and designate the region that first shows abnormal epileptic activity to be the SOZ \citep{Jacobs2012}. However, despite careful planning, sometimes visual analysis of intracranial EEG fails to localize the SOZ clearly \citep{Harroud12}. One crucial reason is that sometimes seizure onsets consist of low amplitude, very fast activity. This activity may not generate appropriate power that can be visually detected until the seizure is well underway. Activity with greater power that can be identified may occur later, by which time seizure activity has spread beyond the actual SOZ and involves brain regions that are involved in seizure occurrence but do not serve as the electrical source. Given that seizures are a directional network phenomenon, our method for mapping directional brain networks (i.e., identifying directional connections) using iEEG data is expected to improve understanding of the brain system and localization of the SOZ.

 iEEG data are high-dimensional multivariate time series recordings of many small regions' neuronal activities at a high temporal resolution (millisecond scale) and spatial resolution (about 10 mm in diameter) and with a strong signal-to-noise ratio (SNR) \citep{Cervenka13}, in contrast to popular functional magnetic resonance imaging (fMRI) with a low temporal resolution and scalp EEG with a low spatial resolution. As such, iEEG data provide valuable information about directional brain networks.

 Mapping directional brain networks based on high-dimensional multivariate time series, however, faces multiple challenges. First, it is difficult to construct a model that can accurately characterize the complex mechanism of a high-dimensional brain system, i.e, how each region's activity depends on many others' activities. Second, the estimation of a high-dimensional model has a large variance. With many regions being studied and enormous possibilities in directional connections among the regions, it is challenging to identify only a few strong connections among enormous candidate ones. Though incorporating anatomic connectivity (AC) information into the directional connectivity model can improve the estimation of directional connections \citep{hahn2019new}, AC information is not always available. Here, we consider mapping directional brain networks without relying on AC information.
 Simple sparsity regularization does not address the challenge because high-dimensional sparse networks have many different forms, most of which do not accurately reflect the brain's functional organization. For example, standard $L_1$-regularized estimates \citep{basu2015regularized,nicholson2017varx} lead to the sparse network in which every region has only a few connections with other regions. However, this sparse network is not consistent with known brain networks in which regions with similar functions tend to be closely connected \citep{petersen2015brain}. Third, the computation for analyzing high-dimensional multivariate time series data can be intensive. Existing approaches to mapping directional networks usually address only a part of these challenges, as explained below.



 Network mapping approaches fall into two major categories: information-theoretic-measure based methods and model-based methods. The former includes correlations, cross-correlations \citep{Kramer08,Schiff05}, cross-coherence \citep{schroder2018fresped}, transfer entropy \citep{vicente2011transfer}, directed transinformation \cite{Hinrichs06}, and directed information \cite{LiuY12}, and many others \citep{vanMierlo13,Wilke11}. Although these measures are fast to compute, they are mainly for quantifying pairwise relationship between regions and ignore system features of the brain in which each region's activity depends on many other regions' activities. Thus, information-measure-based approaches lack the ability to delineate the entire signal pathway of directional connections from regions to regions.

Model-based methods have been developed to describe simultaneous directional connectivity among all the recorded regions. The most popular models include dynamic causal modeling \cite[DCM, ][]{Friston03} and neural mass models \cite[NMM, ][]{David03}, which use ordinary differential equations (ODE) to characterize directional connectivity. Because of their complex mathematical formulation, the DCM and NNM are typically used for low-dimensional brain networks (consisting of only a few brain regions being studied). To address this limitation, \cite{Zhang14DDM,Zhang15BDDM,Zhang17BMODDM} proposed to use linear ODEs to approximate high-dimensional brain systems (consisting of many regions). However, parameter estimation of deterministic ODE models is sensitive to the model specification, data noise, and data-sampling frequency.

We propose to use a state-space multivariate autoregression-based (SSMAR) model for iEEG data to address the limitation of existing methods. First, the state-space framework allows for separating the model error due to the inherent model inadequacy for a complex system and the data measurement error. The SSMAR with the two errors is flexible to approximate different systems and is robust to various deviations from the assumed model. Equally importantly, the formulation of SSMAR is much simpler than ODE models, which thus, enables fast computation for high-dimensional data. 

Different from standard MAR \cite{goebel2003investigating,harrison2003multivariate,korzeniewska2008dynamics} and SSMAR \citep{riera2004state,Cheung10}, our SSMAR is uniquely constructed for analyzing iEEG data to map directional brain networks. It has been widely documented \citep{Newman06,Sporns11} that brain networks have a cluster structure, in which regions are more densely connected with regions in the same cluster than with regions otherwise. Our approach incorporates the cluster structure to greatly improve the model estimation. Specifically, we propose a stochastic blockmodel (SBM)-motivated prior for the SSMAR parameters, restricting the estimated network to have the cluster structure. The SBM \citep{airoldi2008mixed,nowicki2001estimation,durante2014nonparametric,geng2018probabilistic}
is a generative model for the networks in the cluster structure. However, existing applications of the SBM \citep{paul2018random,arroyo2017network} and most cluster identification methods (also called community detection, a terminology often used in social network literature) \citep{goldenberg2010survey,zhao2012consistency} are for observed networks with known edges. The proposed method addresses a more challenging problem of inferring {\it{unobserved networks}} based on multivariate time series measurements of network nodes' activities.


Using the SBM-motivated prior for SSMAR parameters, we develop a Bayesian framework to make inferences about the underlying network. The proposed Bayesian approach has three major advantages. First, our method improves the efficiency in identifying connected brain regions (i.e., a high true positive) and produces scientifically interpretable network results by incorporating the cluster structure into the model. Second, the proposed Bayesian framework accounts for the model error due to the model inadequacy for the complex system as well as the statistical uncertainty in identifying connected regions. Third, the simple SSMAR formulation brings the flexibility to approximate various brain systems and enables fast computation for high-dimensional multivariate time series data. As such, our approach effectively addresses the three major challenges in mapping high-dimensional brain networks.

The rest of the article is organized as follows. In Section 2, we introduce the new SSMAR model for directional brain networks with the cluster structure. We build a Bayesian hierarchical model with an SBM-motivated prior to make inferences of SSMAR parameters and develop an efficient Markov chain Monte Carlo (MCMC) simulation algorithm for the ensuing posterior inference. In Section 3, we apply the developed Bayesian model to data simulated under two different model settings and network patterns and compare the ensuing results with those of existing network mapping methods. We show that the proposed method is robust to various deviations from the assumed model and outperforms existing methods by achieving much higher accuracy in identifying connected brain regions. In Section 4, we analyze real iEEG data from an epileptic patient and examine his brain network changes over the course of seizure development. Section 5 concludes with a discussion.

\section{Dynamic System Models and Bayesian Inference}

\subsection{The State-Space MAR Model}
Let $\bm{y}(t)= (y_1(t), \ldots, y_d(t))'$ be observed iEEG measurements of $d$ brain regions (equivalently $d$ network nodes of the brain network under study) at time $t$ and $\bm{x}(t) = (x_1(t), \ldots, x_d(t))'$ be the underlying neuronal state functions of the $d$ brain regions at time $t$ for $t = 1, \ldots, T$. Since each iEEG electrode directly records one brain region's neuronal activity with a high spatial and temporal resolution, we propose a simple space model that links $y_i(t)$ to $x_i(t)$:
\begin{equation}\label{eq:Obs2}
y_i(t)= c_i\cdot x_i(t) +  \epsilon_i(t), ~~ i = 1, \ldots, d,
\end{equation}
where $c_i$ is a unknown constant, and $ \epsilon_i(t)$ is a data measurement error with mean zero.

For the state model that describes directional connectivity among the $d$ regions at the neuronal level, we propose to use the simplest dynamic system model, i.e., the first-order multivariate-autoregression (MAR), for $\boldsymbol{x}(t)$:
\begin{equation*}\label{eq:MAR}
x_{i}(t) = \sum_{j=1}^{d} A_{ij} \cdot x_{j}(t-1) + \eta_i(t), ~~ i = 1, \ldots, d, ~ t = 1, \ldots, T,
\end{equation*}
where $\eta_i(t)$ is the model error due to the model inadequacy in characterizing the dynamics of region $i$.

 Our goal is to develop a parsimonious model to detect the existence of temporal dependence among neuronal activities of regions rather than building a comprehensive model that can explain all the neuronal activities. Due to the high-dimensionality and the current limited understanding of the brain system, it is extremely difficult to build such a comprehensive dynamic system model. Even though more complex models, such as high-order MARs, may fit the observed data better, they still suffer from the model inadequacy. More seriously, high-order MARs have large estimation errors because they have at least $d^2$ more parameters than first-order MARs. Consequently, the first-order MAR is more efficient for detecting connected regions and addresses our needs.

Under the state-space MAR, identifying connected regions and mapping the brain network are equivalent to selecting statistically significant nonzero $A_{ij}$s. To distinguish nonzero directional connections from zero ones, we introduce indicators for $A_{ij}$s:

\begin{equation}\label{eq:SBSSVAR}
x_{i}(t) = \sum_{j=1}^{d} \gamma_{ij} \cdot A_{ij} \cdot x_{j}(t-1) + \eta_i(t), ~~ i = 1, \ldots, d, ~ t = 1, \ldots, T,
\end{equation}
where $\gamma_{ij}$ is an indicator, taking values either 0 or 1. We use $\gamma_{ij}$s to stand for the set of indicators $\{\gamma_{ij},~i,j=1,\ldots,d\}$. The use of indicators is similar to the ``spike and slab'' prior  \citep{ishwaran2005spike,miller2002subset,theo2004mapping} in the Bayesian variable selection framework \citep{brown1998multivariate,george93,george97,yi2003stochastic}. Under \eqref{eq:SBSSVAR}, identifying connected brain regions, i.e., selecting directional network edges, is equivalent to selecting nonzero $\gamma_{ij}$s, which is the focus of our model estimation.

The observation model \eqref{eq:Obs2} and the state model \eqref{eq:SBSSVAR} together are the proposed state-space MAR (SSMAR) for the brain's directional connectivity. Note that the first-order SSMAR is different from the first-order MAR: The former is robust to violations of model assumptions, but the latter is not. This is because the SSMAR uses two error terms, $\eta_i(t)$ and $\epsilon_i(t)$, to accommodate the model inadequacy and measurement error separately.

We let $\eta_i(t)\stackrel{\mbox{i.i.d}}{\sim} \mbox{N}(0,1)$ for several reasons. First, $c_i$ in (1) and the variance of $\eta_i(t)$ are not uniquely defined. Since we treat the former as unknown, we fix the latter at 1 to avoid the identifiability issue. Second, letting $\eta_i(t)$ be independent between regions enables $\gamma_{ij}$ and $A_{ij}$ to capture the dependence between regions more efficiently than otherwise. Third, letting $\eta_i(t)$ be independent over time brings parsimony to the model. Again, our purpose is to detect the existence of temporal dependence between regions' iEEG rather than capturing all possible temporal dependence. Similarly, for the latter two reasons, we let $\epsilon_i(t)\stackrel{\mbox{i.i.d}}{\sim} \mbox{N}(0,\tau_i)$. We show through simulation studies (Section \ref{sec:Simu}) that our approach is robust to violations of model assumptions.

\subsection{Bayesian Hierarchical Model for SSMAR}\label{sec:Bayes}

Since nonzero $\gamma_{ij}$s define the brain's directional network, we impose the cluster structure on the estimated brain network through using a stochastic blockmodel (SBM)-motivated  \citep{fienberg1985statistical,airoldi2008mixed,nowicki2001estimation,durante2014nonparametric}
prior for $\gamma_{ij}$s. The cluster structure means that regions within the same cluster connect more closely with each other than with regions in a different cluster. The cluster structure fits the brain's functional organization reported in the literature \citep{Newman06,Sporns11} and is also useful in epilepsy diagnosis. For example, regions in the SOZ's cluster are those affected by the SOZ's activities most. Information about the SOZ's cluster and its changes during seizure development can help neurologists assess the effect of seizures on brain functions. In summary, developing the SBM-motivated prior for SSMAR parameters to impose the cluster structure on estimated networks is another important novelty of our approach.

Let $K$ be the pre-specified number of clusters. Let $\bm{m}_i=(m_{i1},\ldots, m_{iK})^\prime$ be a $K$-dimensional vector with only one element being 1 and the rest being 0; $\bm{m}_i$ labels the cluster of region $i$, i.e., $m_{ik}=1$ indicates region $i$ in the $k$th cluster. Let $B_{k_1k_2}$, $k_1,k_2=1,\ldots,K$, denote the prior probability of a nonzero directional connection from a region in cluster $k_2$ to another region in cluster $k_1$. Let $\mathbf{B}$ be a $K\times K$ matrix with entries $B_{k_1k_2}$ for $k_1,k_2=1,\ldots,K$.

\noindent{\textbf{Prior specification for the cluster structure.}} The prior for the brain network with the cluster structure is a joint distribution for indicators $\gamma_{ij}$s, the cluster labels $\bm{m}_i$s, and the probability matrix $\mathbf{B}$ as follows:\vspace{-0.1cm}
\begin{eqnarray}
&\gamma_{ij}|\bm{m}_i,\bm{m}_j,\mathbf{B}\stackrel{\mbox{ind}}{\sim} \mbox{Bernoulli}(\bm{m}_i^\prime~ \mathbf{B}~ \bm{m}_j);\label{eq:priorIndicator}\\
& \bm{m}_i\stackrel{\mbox{i.i.d}}{\sim} \mbox{Multinomial}(1;p_1,\ldots,p_K) ~\mbox{for}~ i=1,\ldots, d,~\mbox{and}~
(p_1,\ldots,p_K)\sim \mbox{Dirichlet}(\bm{\alpha});\label{eq:Multi-Dirichlet}\\
&B_{kk} \stackrel{\mbox{i.i.d}}{\sim} \mbox{Uniform}(l_0, 1)~\mbox{and}~
B_{k_1k_2} \stackrel{\mbox{i.i.d}}{\sim} \mbox{Uniform}(0, u_0), k_1,k_2 = 1, \ldots , K,~ k_1 \neq k_2; \label{eq:priorProb}\vspace{-0.1cm}
\end{eqnarray}
where $l_0$ and $u_0$ are given constants between 0 and 1, and $\bm{\alpha}=(1,\ldots, 1)$, assigning uniform weights to different clusters. The distribution \eqref{eq:priorIndicator} specifies the probabilities of both within-cluster and between-cluster connections. For example, if $m_{ik_1}=1$ and $m_{jk_2}=1$, then $\bm{m}_i^\prime~ \mathbf{B}~ \bm{m}_j={B}_{k_1k_2}$, which is the probability of existing a directional connection from cluster $k_2$ to cluster $k_1$; if $m_{ik}=1$ and $m_{jk}=1$, $\bm{m}_i^\prime~ \mathbf{B}~ \bm{m}_j={B}_{kk}$, which is the prior probability of existing a directional connection between two regions in the same cluster $k$. Since within-cluster connections are dense and strong, while between-cluster connections are sparse \citep{ParkFriston2013}, we let $u_0=0.1$ and $l_0=0.9$. The large difference between $u_0$ and $l_0$ facilitates differentiating within-cluster connections from between-cluster ones and identifying clusters.

The distributions \eqref{eq:priorIndicator}, \eqref{eq:Multi-Dirichlet}, and \eqref{eq:priorProb} together define the SBM-motivated prior for $\gamma_{ij}$s. Our goal is to identify clusters and select significant edges by estimating the cluster labels for regions, $\bm{m_i}$s, and the indicators for edges, $\gamma_{ij}$s.

\noindent{\textbf{Prior specification for $A_{ij}$s.}} We assign a normal prior to $A_{ij}$:
\begin{equation}\label{eq:PriorA}
A_{ij}\stackrel{\mbox{i.i.d}}{\sim}  \mbox{N}(0, \xi_0^2),
\end{equation}
where $\xi_0 $ is a positive constant so that the density of $A_{ij}$ is almost flat within its domain.

\noindent{\textbf{Priors for other parameters.}} Let $\bm{x}(0)=(x_1(0),\ldots,x_d(0)'$, $\bm{c} = (c_1, \ldots, c_d)'$, $\bm{\mu}=(\mu_1,\ldots,\mu_d)'$, and $\bm{\tau}=(\tau_1,\ldots,\tau_d)'$. We assign the following priors to the rest parameters:
\begin{eqnarray}\label{eq:PriorPara}
 x_i(0)\stackrel{\mbox{ind}}{\sim}\mbox{N}(\mu_i,1), ~\mu_i\stackrel{\mbox{i.i.d}}{\sim}\mbox{N}(0,\xi^2_1),~
 c_i\stackrel{\mbox{i.i.d}}{\sim}\mbox{N}(0,\xi^2_1),~
 p(\tau_i)\propto \frac{1}{\tau_i^{1+\rho_0}}\exp\{-\frac{\rho_0}{\tau_i}\},~ i=1,\ldots,d,
 \end{eqnarray} where $\rho_0$ is a pre-specified small positive constant to give an almost flat prior for $\bm{\tau}$ and $\xi_1$ is a large positive constant to give almost flat priors for $c_i$ and $\mu_i$.

\noindent\textbf{Joint posterior distribution.} All the parameters to be estimated in the proposed Bayesian framework are
$\bm{\Theta} = \{\bm{\Gamma},\mathbf{B},\mathbf{M},\mathbf{A},  \bm{c}, \bm{\tau}, \bm{\mu},\bm{p} \}$, where $\bm{\Gamma}$ is a $d\times d$ matrix with entries $\gamma_{ij}$ for $i, j = 1, \ldots, d$, $\mathbf{M}$ is a $K\times d$ matrix with the $i$th column being $\bm{m}_i$, $\mathbf{A}$ is a $d\times d$ matrix with entries $A_{ij}$ for $i, j = 1, \ldots, d$, and $\bm{p}=\{p_1,\ldots,p_K\}$.

Let $\mathbf{X}=\{\bm{x}(0),\ldots, \bm{x}(T)\}$ and $\mathbf{Y}=\{\bm{y}(1),\ldots, \bm{x}(T)\}$. The SSMAR model \eqref{eq:Obs2} and \eqref{eq:SBSSVAR} with prior distributions \eqref{eq:priorIndicator}, \eqref{eq:Multi-Dirichlet}, \eqref{eq:priorProb}, \eqref{eq:PriorA}, and \eqref{eq:PriorPara} lead to the posterior distribution:
$p(\mathbf{X}, \bm{\Theta} | \mathbf{Y}) \propto p(\mathbf{Y}|\mathbf{X},\bm{\Theta})\cdot p(\mathbf{X}|\bm{\Theta})\cdot p(\bm{\Theta}).$
The detailed formulation of the joint posterior distribution is provided in the Appendix.

\subsection{EM Algorithm for Setting Initial Values and Hyperparameter}
We simulate from $p(\mathbf{X}, \bm{\Theta} | \mathbf{Y})$ with a partially collapsed Gibbs Sampler \cite[]{van08}, whose Markov Chain Monte Carlo (MCMC) simulation steps are provided in the Appendix.

The MCMC simulation can take many iterations to converge especially for large $d$. To address this issue, following the practice suggested in \cite[Chapter 13.1,][]{gelman2013bayesian}, we use an expectation-maximization (EM) algorithm to find the starting values for the MCMC simulation. Specifically, we optimize $p(\mathbf{Y}|\hat{\bm{\Theta}})=\int p(\mathbf{Y}|\hat{\bm{\Theta}},\mathbf{X})\cdot p(\mathbf{X}|\bm{\Theta})d\mathbf{X}$ by the EM algorithm, in which the state functions $\mathbf{X}$ are treated as missing values. The output of the EM algorithm, $\hat{\bm{\Theta}}$ in the final step, is used as the initial value for the following 10,000 MCMC iterations. For all our simulation and real data analysis, we verified that the MCMC algorithm converged upon evaluating the Gelman-Rubin statistic \cite{Gelman1992}.

We need to determine the value of $K$, the number of clusters, for the proposed Bayesian model. Standard approaches to selecting hyperparameters for Bayesian methods include information criteria and cross-validation. However, these methods are time-consuming for large $d$, because they all require running the posterior simulation for each candidate $K$. We propose to select the value for $K$ by the EM algorithm. Specifically, we let $K=d$ in our EM algorithm. We set the initial values of $m_{ii}$ to 1 for $i=1,\ldots,d$, that is, we let each region form one independent cluster at the start of the EM algorithm. As the algorithm iterates, several regions fall into the same cluster, and the number of distinct clusters of the $d$ regions becomes stable. Since the EM algorithm can find the number of clusters that leads to a locally optimal posterior, we let the $K$ in the Bayesian model be the number of distinct clusters in the final step of the algorithm.


\subsection{Posterior Inference}
We use two posterior probabilities to map the brain network: $\hat{P}_{ij}^{m} = \frac{1}{S} \sum_{s=1}^S \delta(\bm{m}_i^{(s)}, \bm{m}_j^{(s)})$ and $\hat{P}_{ij}^{\gamma}= \frac{1}{S} \sum_{s=1}^S \gamma_{ij}^{(s)}$, where $S$ is the total number of MCMC samples after burn-in. The former, called the clustering probability, is the posterior probability of two regions $i$ and $j$ in the same cluster; and the latter, called the network edge probability, is the posterior probability of nonzero directional connectivity from region $j$ to $i$.
We use $\hat{P}_{ij}^{m}$, $i,j=1,\ldots,d$, to identify clusters. Given a threshold $\hbar^m$, if  $\hat{P}_{ij}^{m}>\hbar^m$, regions $i$ and $j$ are put in the same cluster; if additionally, $\hat{P}_{jk}^{m}>\hbar^m$, then the three regions $i$, $j$, and $k$ are put in the same cluster regardless of the value of $\hat{P}_{ik}^{m}$. We use $\hat{P}^\gamma_{ij}$ to select directional network edges. Given a threshold $\hbar^\gamma$, if $\hat{P}_{ij}^{\gamma}>\hbar^\gamma$, we deem the directional connection from region $j$ to $i$ nonzero and select the directional network edge from $j$ to $i$.

\noindent\textbf{Choice of thresholds.} The total numbers of potential network edges and possible network patterns are enormous for high-dimensional networks. Because of the uncertainty resulted from the high-dimensionality, posterior probabilities $\hat{P}^m_{ij}$ and $\hat{P}^\gamma_{ij}$ are all small. To address this issue, many Bayesian methods select variables based on the ranks of their posterior probabilities \citep{Zhang14Bayes,Zhang15BDDM}. We here propose to determine the thresholds for $\hat{P}^m_{ij}$ and $\hat{P}^\gamma_{ij}$ based on their significance/p-values under the null hypothesis that all the regions are \textit{independent} from each other, as explained in detail below.

We first generate a null data set $\mathbf{Y}^0$ that satisfies the null hypothesis. Specifically, given long iEEG time series before seizure onsets, we randomly sample a short segment ${Y}^0_i=\{y_i(t), t=t_i+1,\ldots,t_i+T\}$ of each region $i$ and let the pairwise distance between any two regions' segments, $|t_i-t_j|$, no smaller than $2T$. All the regions' segments ${Y}^0_i$, $i=1,\ldots,d$, form $\mathbf{Y}^0$, in which the temporal dependence of each region's time-series data points remains while the dependence between regions' time series is almost none. Applying our Bayesian method to $\mathbf{Y}^0$, we obtain the ensuing the clustering probabilities and network edge probabilities, which form the empirical null distributions for $\hat{P}^m_{ij}$s and $\hat{P}^\gamma_{ij}$s, respectively. We evaluate the p-values of $\hat{P}^m_{ij}$s and $\hat{P}^\gamma_{ij}$s based on the null distributions and determine the thresholds for $\hat{P}^m_{ij}$s and $\hat{P}^\gamma_{ij}$s corresponding to the chosen p-value. We here use the p-value of 1\% to ensure a low false positive rate.
\section{Simulation Study}\label{sec:Simu}
\subsection{Example 1: Simulation from A Third-Order SSMAR}
\begin{figure}
\centering
\subfigure[Simulated Cluster Structure]
{\label{fig:TrueNetwork1}
     \includegraphics[width=5.1cm,height=4.0cm,trim= 20mm 30mm 10mm 30mm,clip=TRUE]{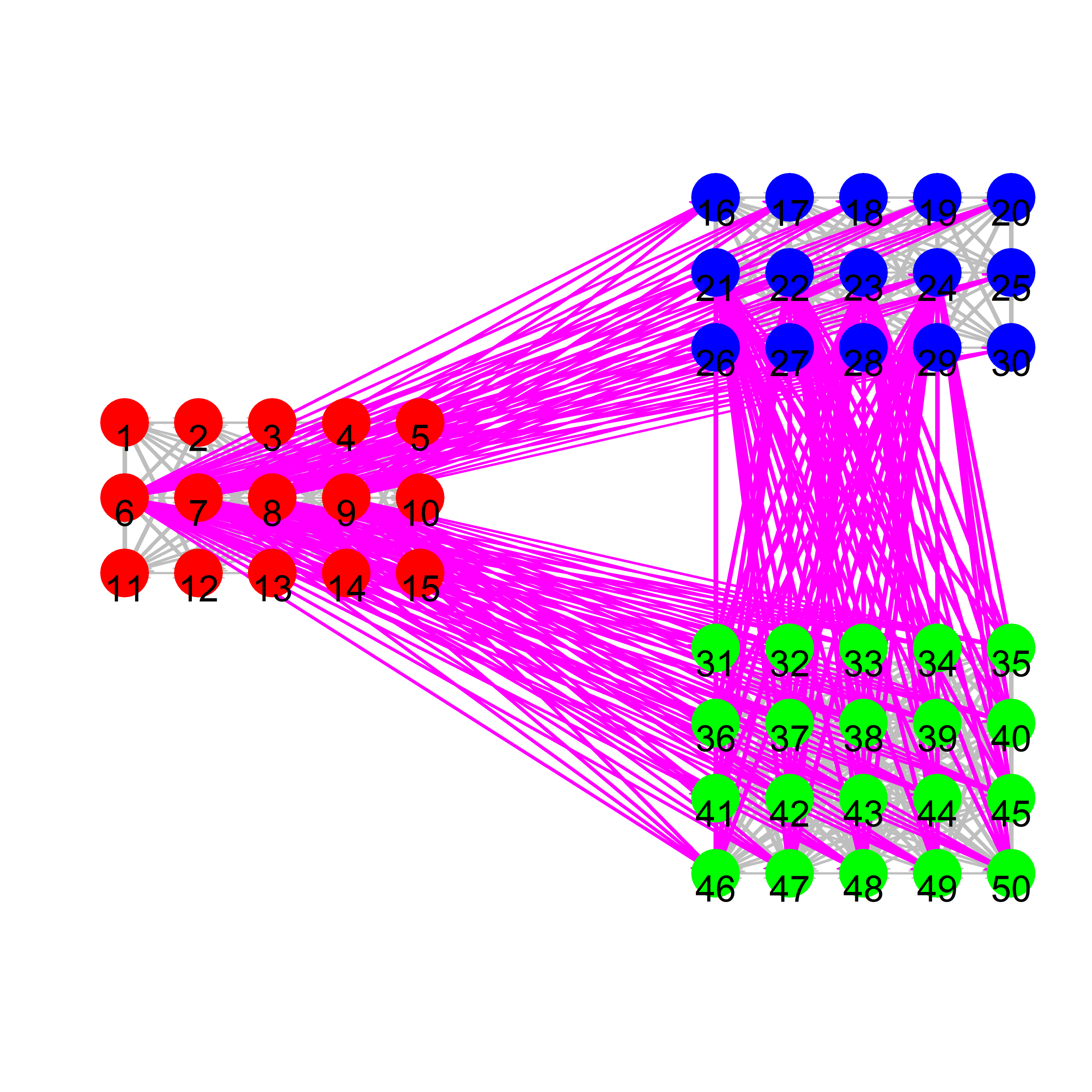}
     }
  \subfigure[ROC Curves]{\label{fig:ROC}
    \includegraphics[height=4.0cm,width=5.1cm,trim= 0mm 0mm 5mm 5mm,clip=TRUE]{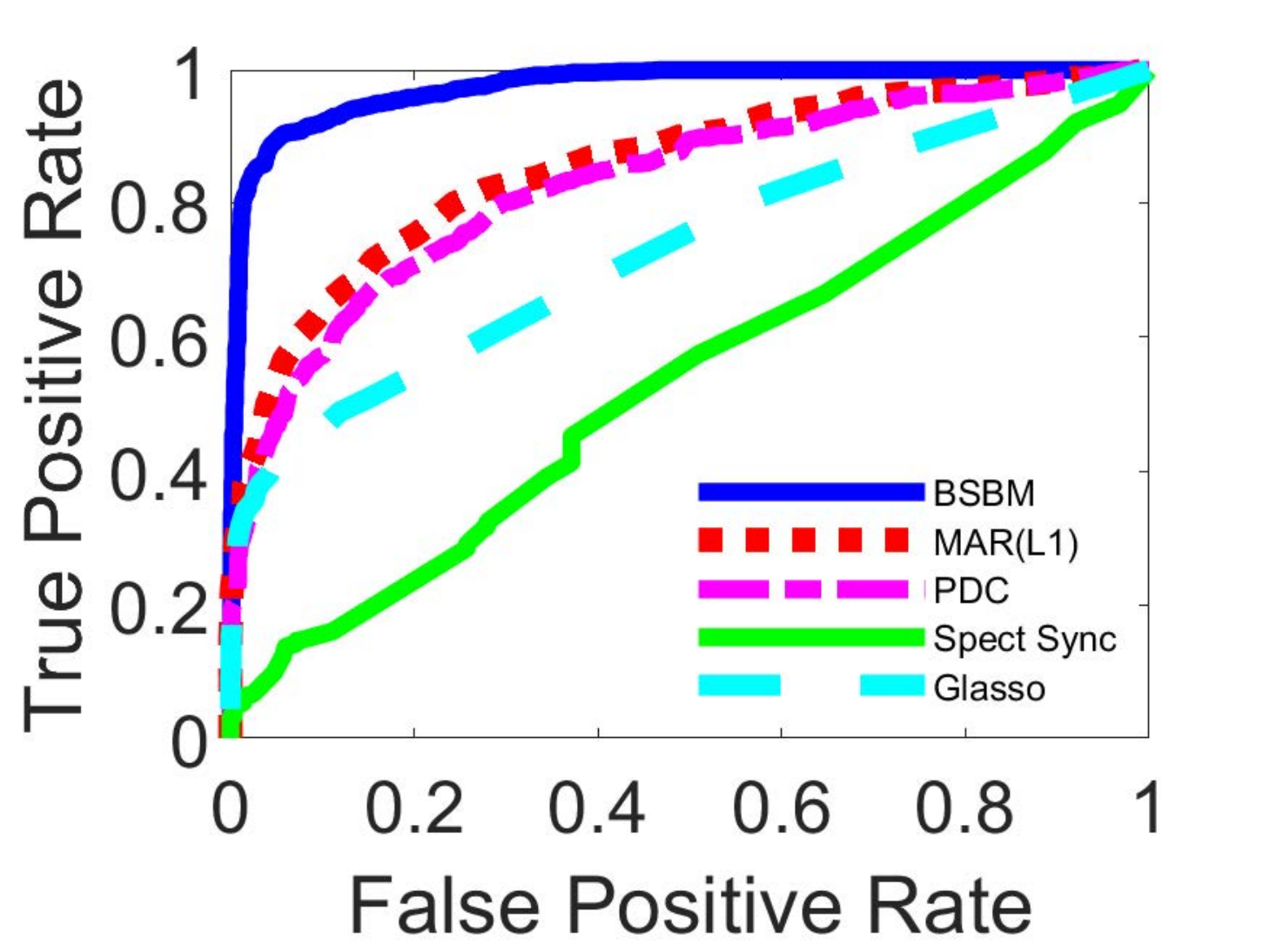}
    }
     \subfigure[Estimated Brain Network]{\label{fig:EstNetwork1}
    \includegraphics[height=4.0cm,width=5.1cm,trim= 20mm 30mm 10mm 30mm,clip=TRUE]{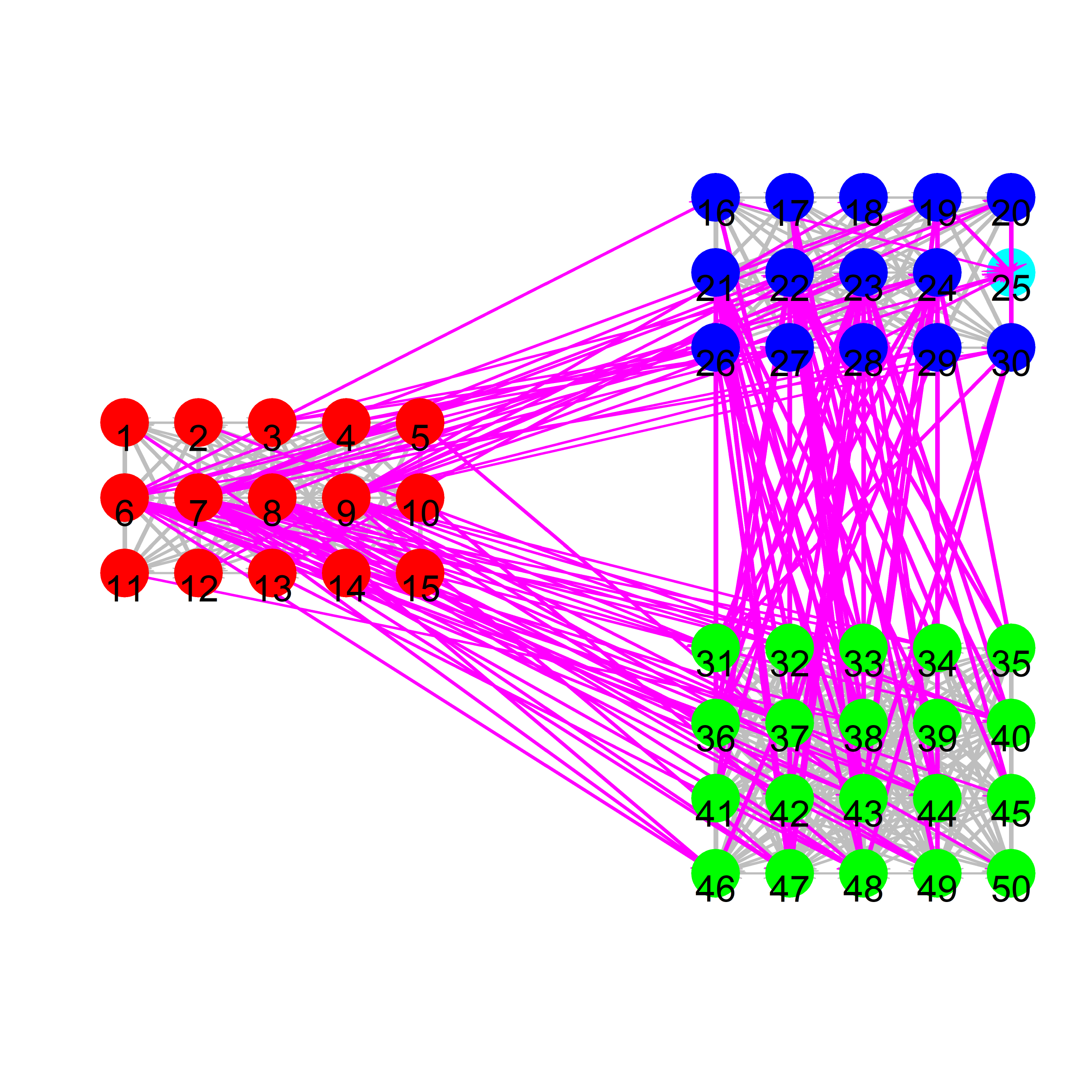}
    }
    \caption{(a) The true simulated network structure. (b) The ROC curves of the proposed Bayesian method with a SBM-motivated prior (BSBM) and competing methods including MAR($L_1$), PDC, the spectrum synchronicity, and Glasso. (c) The estimated network corresponding to 1\% p-value.}
\end{figure}

We simulated multivariate time-series data from the following third-order SSMAR.
\begin{eqnarray*}
x_i(t)& =& \sum_{j=1}^{d} A_{1,ij}~ x_j(t-1) + \sum_{j=1}^{d} A_{2,ij}~x_j(t-2) + \sum_{j=1}^{d} A_{3,ij}~ x_j(t-3) + \eta_i(t)~\mbox{and}\\
y_i(t) &=& c_i \cdot x_i(t) + \epsilon_i(t).
\end{eqnarray*}
 The above system has three clusters of size 15, 15 and 20. We consider region $j$ has a directional influence over $i$, if at least one of $A_{1,ij}$, $A_{2,ij}$, and $A_{3,ij}$ is nonzero. Figure \ref{fig:TrueNetwork1} shows the simulated network pattern, where the presence of a directional connection is indicated by an edge (grey edges for within-cluster connections and purple edges for between-cluster connections).

We simulated $\eta_i(t)$ from the model
\begin{equation}\label{eq:eta}
\bm{\eta}(t)=0.5\bm{\eta}(t-1)+\bm{\delta}(t)~\mbox{and}~ \bm{\delta}(t)\stackrel{i.i.d}{\sim}\mbox{MNV}(0,\bm{\Sigma}_1),
 \end{equation} where $\bm{\Sigma}_1$ is a block diagonal matrix with each block corresponding to one cluster. The diagonal entries of $\bm{\Sigma}_1$ all equal 1 and off-diagonal entries in diagonal submatrices follow Uniform(0,0.5). The upper bound of off-diagonal entries is chosen such that $\bm{\Sigma}_1$ is strictly positive definite. 

We generated the observation errors $\boldsymbol{\epsilon}(t)=(\epsilon_1(t),\ldots, \epsilon_d(t))^\prime$ from the model
\begin{equation}\label{eq:epsilon}
\bm{\epsilon}(t) = 0.5\bm{\epsilon}(t-1) + \bm{\zeta}(t)~\mbox{and}~\bm{\zeta}(t)\stackrel{i.i.d}{\sim}\mbox{MVN}(0, \mathbf{D}^{\frac{1}{2}}\bm{\Sigma}_2\mathbf{D}^{\frac{1}{2}}),\end{equation}where $\bm{\Sigma}_2$ is created in the same way as $\bm{\Sigma}_1$, and $\mathbf{D}$ is a $d$-by-$d$ diagonal matrix with the diagonal entries chosen such that the SNRs of all the time series equal 10. The median SNR of real iEEG data is much higher than 10 \citep{Zhang14DDM}. As such, the simulated model errors and data errors are all spatially and temporally correlated, which violates the model assumptions of the proposed SSMAR.

 Using the simulated edges as the true values, we calculated false positive rates (FPR) and true positive rates (TPR) of network edge selection based on different thresholds for $\hat{P}^\gamma_{ij}$s. For comparison, we examined the FPRs and TPRs of popular competing methods, including the third-order MAR with $L_1$ regularization (implemented by using the R package BigVAR \citep{nicholson2017varx}), denoted by MAR($L_1$), partial directed coherence (PDC) \citep{baccala2001partial}, the spectrum synchronicity \citep{euan2018spectral}, and graphical lasso (Glasso) \citep{friedman2014glasso,witten2011new}.  Figure \ref{fig:ROC} shows the ROC curves of TPRs vs. FPRs for these methods. The proposed Bayesian method with the SBM-motivated prior (BSBM) outperformed the other methods as evidenced by its much greater TPRs given the same FPRs.

 Figure \ref{fig:EstNetwork1} shows the estimated network pattern using the thresholds corresponding to 1\% p-value for $\hat{P}^m_{ij}$ and $\hat{P}^\gamma_{ij}$. The proposed method was able to identify three clusters. For detecting the directional connections among the 50 regions, the overall TPR and FPR are 0.84 and 0.02. More specifically, the TPR and FPR are 0.95 and 0 for within-cluster connections  and 0.45 and 0.02 for between-cluster connections. The comparably low TPR for selecting between-cluster connections is due to several reasons. First, since the clustering is subjective, our selection of directional network edges based on $P^\gamma_{ij}$ does not account for the identified clusters. As within-cluster connections (accounting for 32.6\% of all candidate connections) are much denser than between-cluster connections (9.0\% of all candidate connections), network edge selection is more towards selecting within-cluster connections, so that the overall network edge selection accuracy is high. Second, the number of candidate between-cluster connections is enormous and even more than the total number of true network edges. As such, the true between-cluster connections are highly sparse and more difficult to identify than within-cluster connections. Third, since the number of null connections is large, we used a high threshold for $P^\gamma_{ij}$ to avoid many false selections, which also leads to a low TPR for selecting between-cluster connections. Overall, the proposed method outperformed existing methods by achieving a higher TPR and an almost zero FPR.

 In summary, this simulation demonstrates the robustness of our SSMAR to violations of model assumptions and its efficiency in identifying connected regions and clusters.

\subsection{Example 2: Simulation from the Dynamic Causal Modeling}
\begin{figure}[h]\vspace{-0.4cm}
\centering
\subfigure[\footnotesize{Simulated Network}] 
{    \label{fig:DCMfMRINetwork}
    \includegraphics[width=5cm,height=3.90cm,trim= 32mm 30mm 5mm 20mm,clip=TRUE]{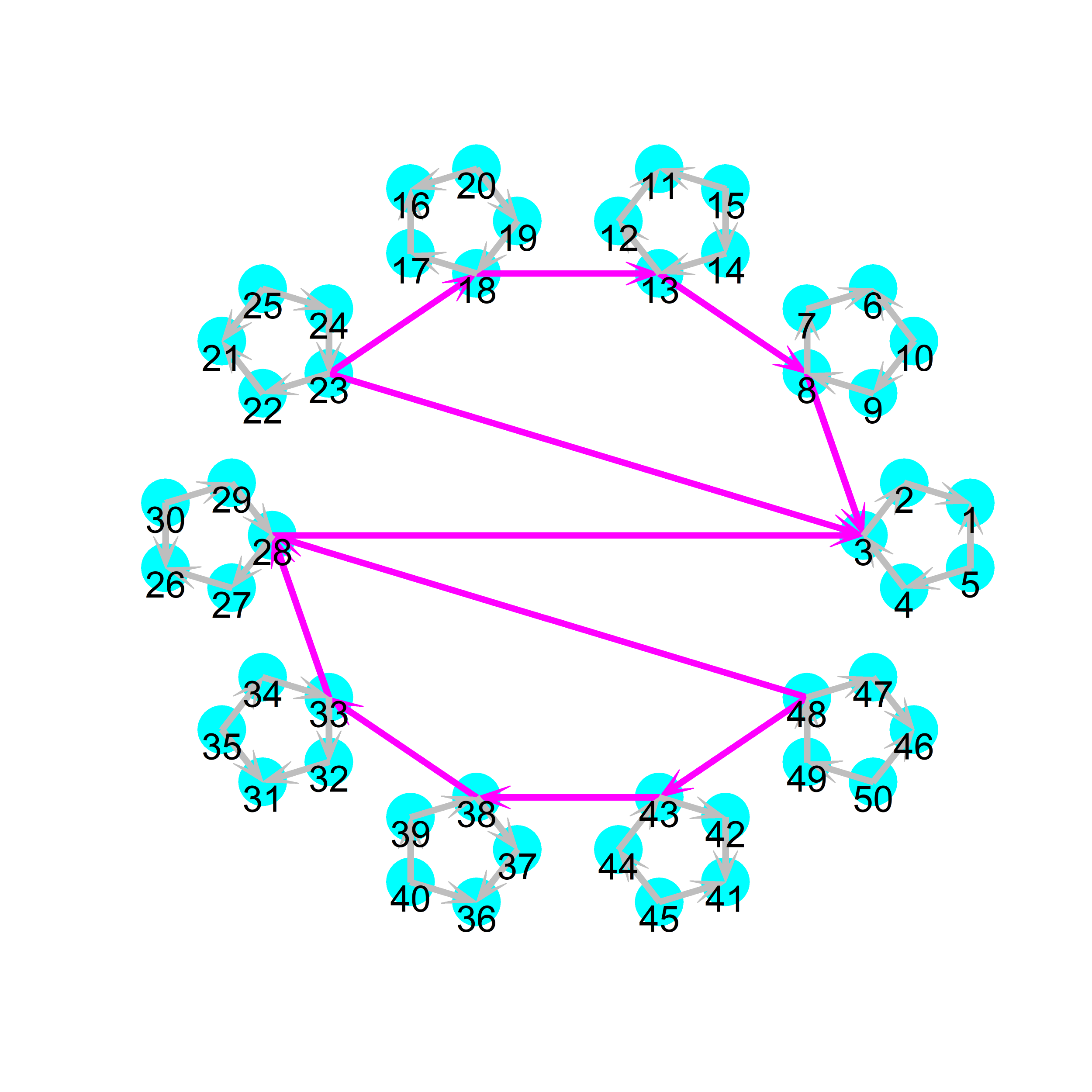}
}
\subfigure[\footnotesize{ROC Curves for Data with 2714 Time Points}] 
{    \label{fig:ROCDCMfMRI1000}
    \includegraphics[width=5cm,height=3.90cm,trim= 0mm 0mm 12mm 02mm,clip=TRUE]{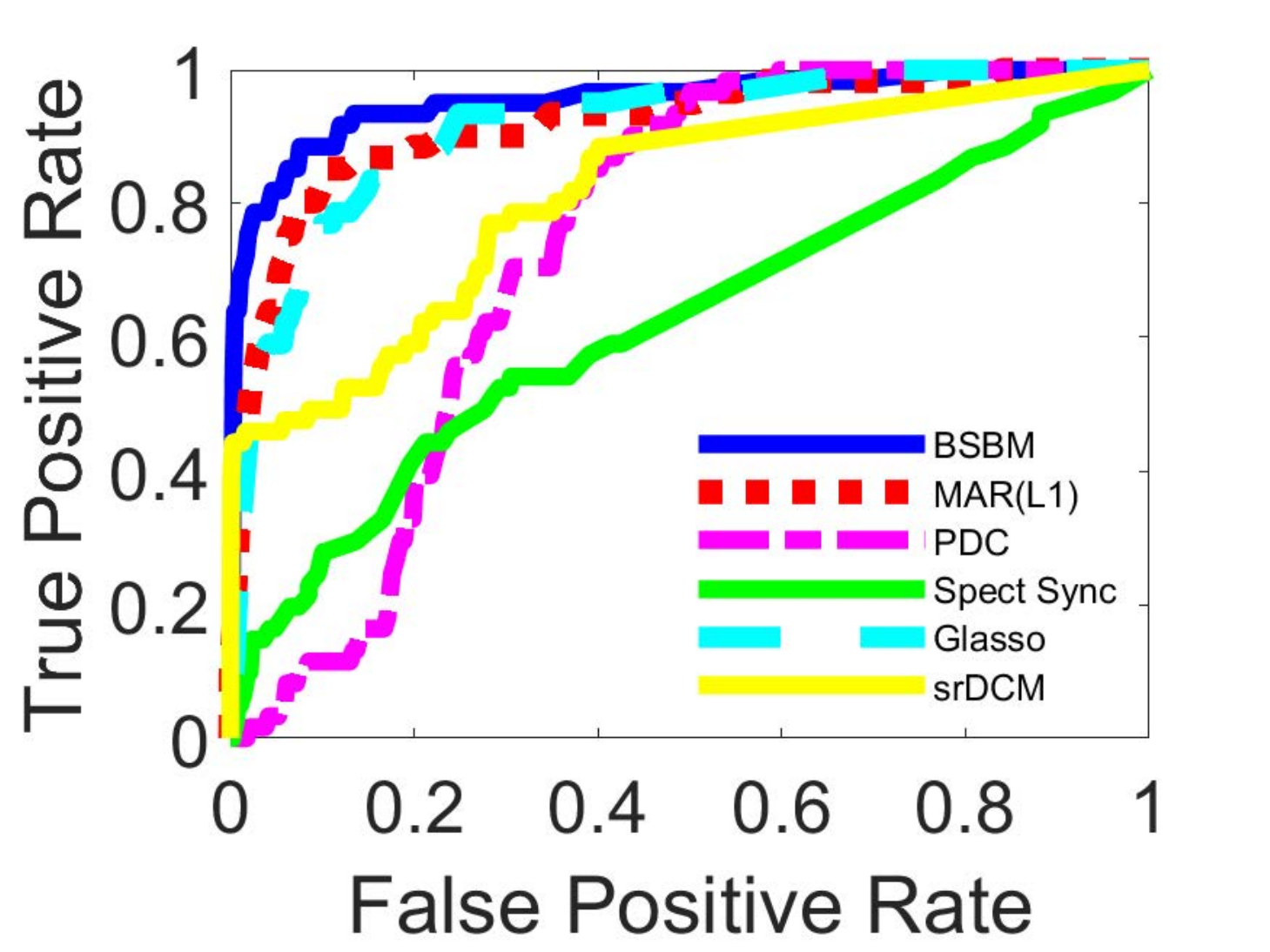}
}
 \subfigure[\footnotesize{ROC Curves for Data with 1000 Time Points}]
{
    \label{fig:ROCDCMfMRI500}
    \includegraphics[width=5cm,height=3.90cm,trim= 0mm 0mm 12mm 02mm,clip=TRUE]{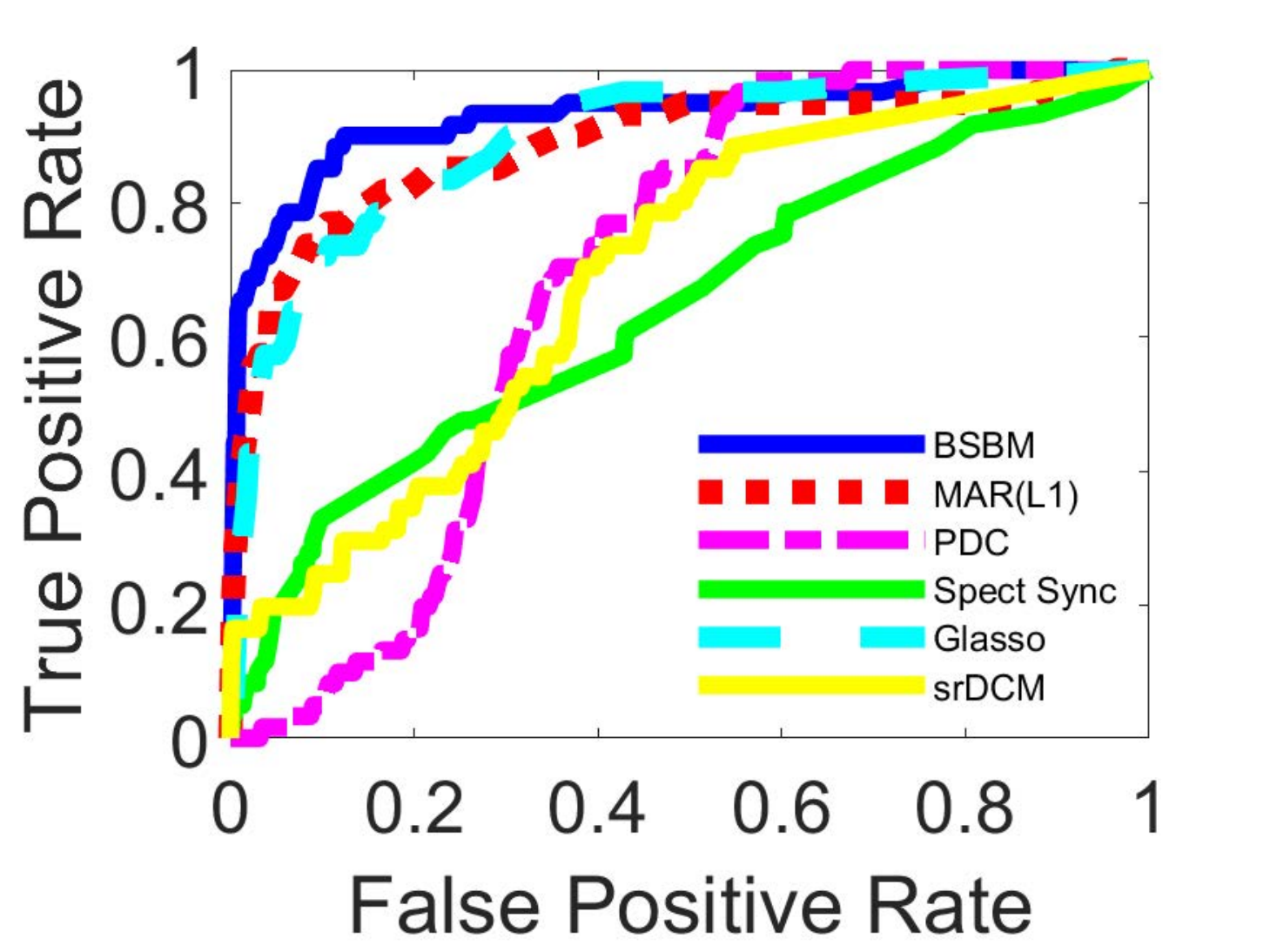}
}
\vspace{-0.42cm}\caption{\small{Simulation studies of two generative models for fMRI data}. \label{fig:DCMfMRI1}\footnotesize{(a) The simulated (true) network pattern. (b) ROC curves of network edges selection for the simulated data at 2714 time points. (c) ROC curves of network edge selection for the simulated data at 1000 time points.}}\vspace{-0.3cm}
\end{figure}

We simulated time series from a 50-dimension dynamic system given by the dynamic causal modeling (DCM) \cite{Friston03}, the most popular ODE-based model for the brain's directional connectivity. The DCM is for low-dimensional brain networks. We expanded its state model to be high-dimensional and the same as that of the sparse regression-DCM  (srDCM) \citep{frassle2018generative}, an extension of the DCM for high-dimensional brain networks. We used this high-dimensional state model to generate $\mathbf{x}(t)$ of 50 regions. Then we simulated $\mathbf{y}(t)$ based on the observation model of the DCM, which describes the transformation of neuronal activity $\mathbf{x}(t)$ into observed $\mathbf{y}(t)$. The signal-to-noise was set to be 1, which was considered small in the literature \citep{frassle2018generative}. Figure \ref{fig:DCMfMRINetwork} shows the simulated directional network among 50 regions.

We applied the proposed BSBM to simulated $\mathbf{y}(t)$ with 2714 time points, which were identical to those of the simulated data under the srDCM \cite{frassle2018generative}. We also applied the BSBM to down-sampled data with 1000 time points. Figures \ref{fig:ROCDCMfMRI1000} and \ref{fig:ROCDCMfMRI500} show the ROC curves of the BSBM and other competing methods for the data of two frequencies. We also analyzed the simulated data by the srDCM. Though the proposed model is distinct from the DCM and srDCM, our method was robust to  model specification, data noise, and data-sampling frequency and outperformed existing methods by achieving the largest area under the ROC curve.

\section{Real iEEG Data Analysis}

We applied the proposed method to iEEG data of an epileptic patient, who had 64 electrodes placed on the exposed surface of his brain, as shown in Figure \ref{fig:Sub1}. iEEG recorded the patient's brain activities in 3 seizures. The sampling rate of this patient's iEEG data was 4000 Hz. We down-sampled the iEEG data to 1000 Hz, a typical rate used in the literature \citep{Burns14,Zhang14DDM}. EEG experts manually examined the data and determined seizure onset times and the SOZ, which was G37. A responsive neurostimulation system was later implanted in his brain with a lead placed on G37. The use of RNS has significantly reduced his seizure occurrences. This confirms that the SOZ was accurately located. In our analysis, we treated seizure onset times as given, since the detection of seizure onset time is not difficult. However, we did not use the location information of the SOZ when mapping the directional brain network among recorded brain sites. The SOZ was treated as unknown and equally as other brain sites. As such, we could validate our network results against the location information of the SOZ.

Channels 63 and 64, as the reference electrodes, were removed from the analysis. We evaluated connectivity among the rest 62 regions. To minimize the residual artifacts of 60 Hz electrical noise, we used a 60 Hz notch filter during the primary recording and removed the first principal component through the principal component analysis.

Once a seizure starts, the connection strength between the SOZ and other regions increases  \citep{englot2016regional}, resulting in abnormally synchronized or excessive neuronal activities in other regions \citep{fisher2014ilae}. Thus, an effective brain network mapping methods should reveal different brain networks before and after seizure onset: More regions are expected to be affected by the activities from the SOZ after the seizure onset. We applied our method to map brain networks in the periods around the seizure onset time and examined the effectiveness of our method in revealing different brain networks before and after seizure onset. We focused on four time periods:  26 to 50 seconds before seizure onset, 1 to 25 seconds before seizure onset, 1 to 25 seconds after seizure onset, and 26 to 50 seconds after seizure onset. To ensure effective approximation of the underlying complex brain system by the SSMAR and also to accommodate potential variation of brain activities over time, we applied the developed method to each 1-second iEEG segment (containing 1000 time series measurements) independently. In total, we analyzed 300 1-second iEEG data segments (4 periods $\times$ 25 seconds $\times$ 3 seizures).

For each 1-second data segment and for each pair of regions $i$ and $j$, we obtained their clustering probability $\hat{P}^m_{ij}$ and network edge probabilities $\hat{P}^\gamma_{ij}$ and $\hat{P}^\gamma_{ji}$. For each seizure period, we took average of posterior probabilities in 75 segments and denoted the ensuing average posterior probabilities by $\bar{P}^m_{ij}$, $\bar{P}^\gamma_{ij}$ and $\bar{P}^\gamma_{ji}$. We identified clusters and connected brain regions and mapped brain networks for four seizure periods based on these average probabilities. This analysis is consistent with the medical practice where reliable epilepsy diagnosis is based on combined information of iEEG recordings of at least 3 seizures \citep{marks1998semiology}.

\begin{figure}[h]
\centering
\subfigure[\footnotesize{$t\in [-50 ,-25]$ seconds}] 
{    \label{fig:before25_2}
    \includegraphics[width=0.23\textwidth,height=4.0cm,trim= 70mm 67mm 60mm 22mm,clip=TRUE]{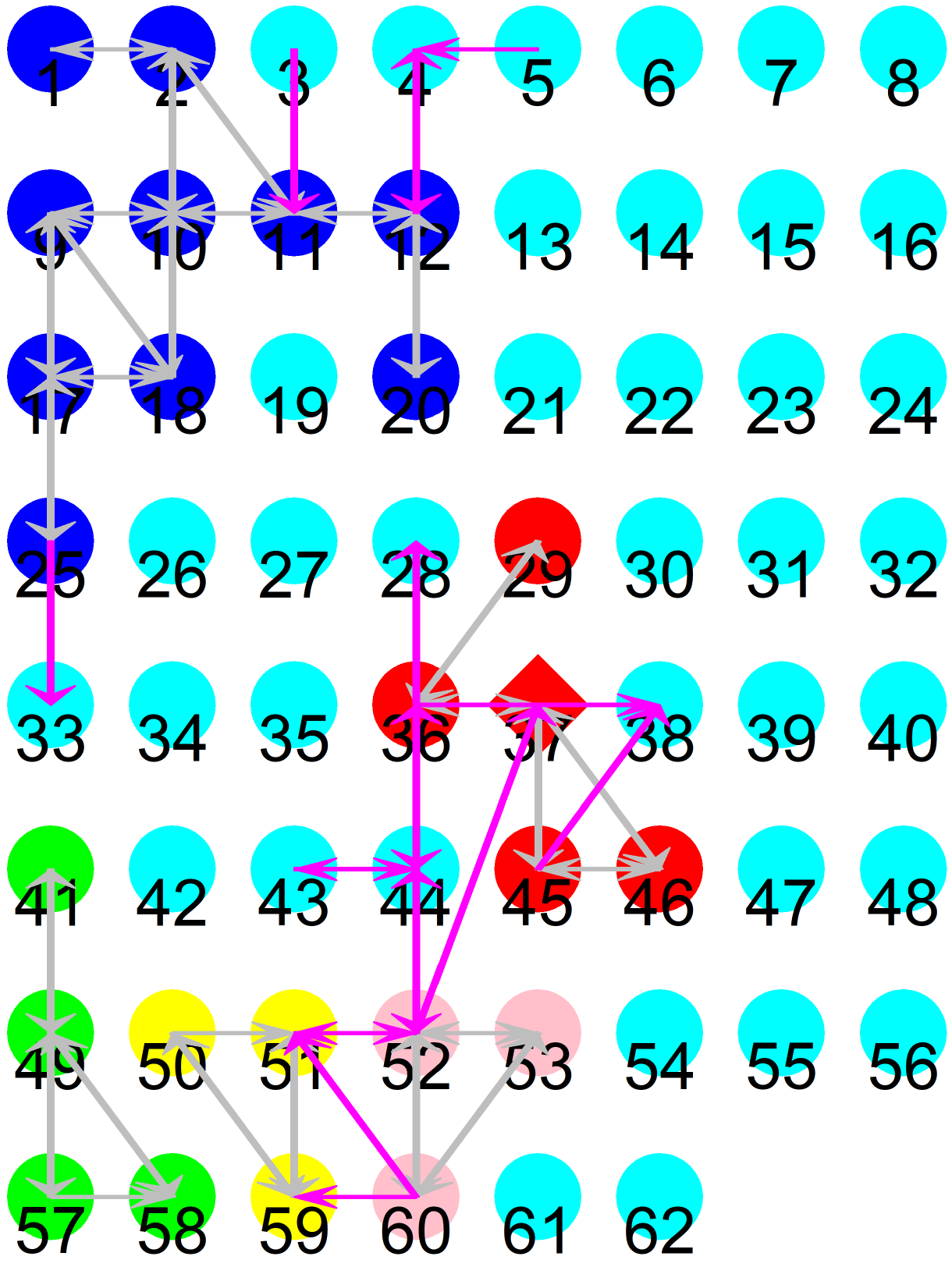}
}
\subfigure[\footnotesize{$t\in [-25 ,0]$ seconds}] 
{    \label{fig:before25_1}
    \includegraphics[width=0.23\textwidth,height=4.0cm,trim= 70mm 67mm 60mm 22mm,clip=TRUE]{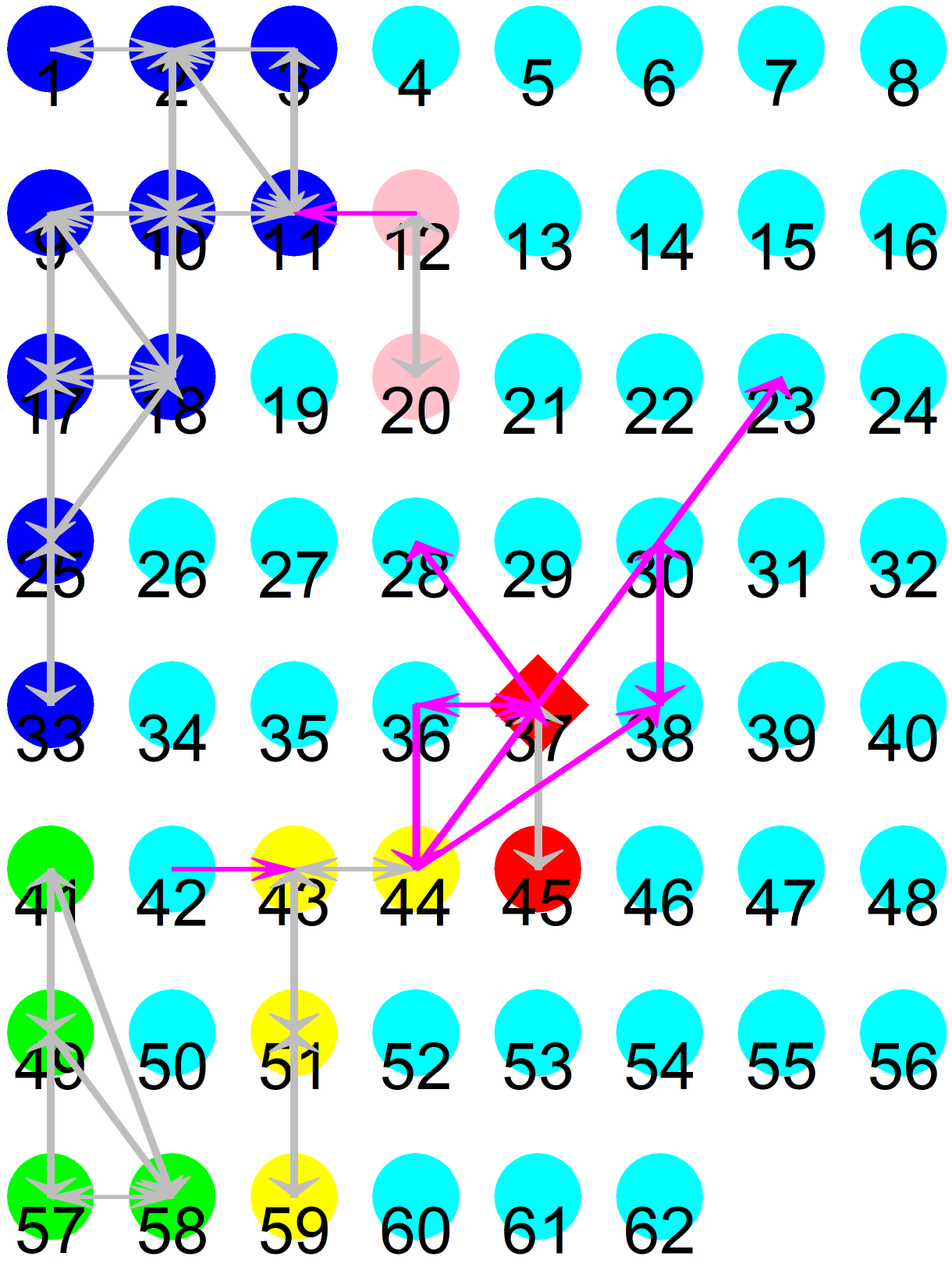}
}
 \subfigure[\footnotesize{$t\in [0 ,25]$ seconds}]
{
    \label{fig:after1_25}
    \includegraphics[width=0.23\textwidth,height=4.0cm,trim= 70mm 67mm 60mm 22mm,clip=TRUE]{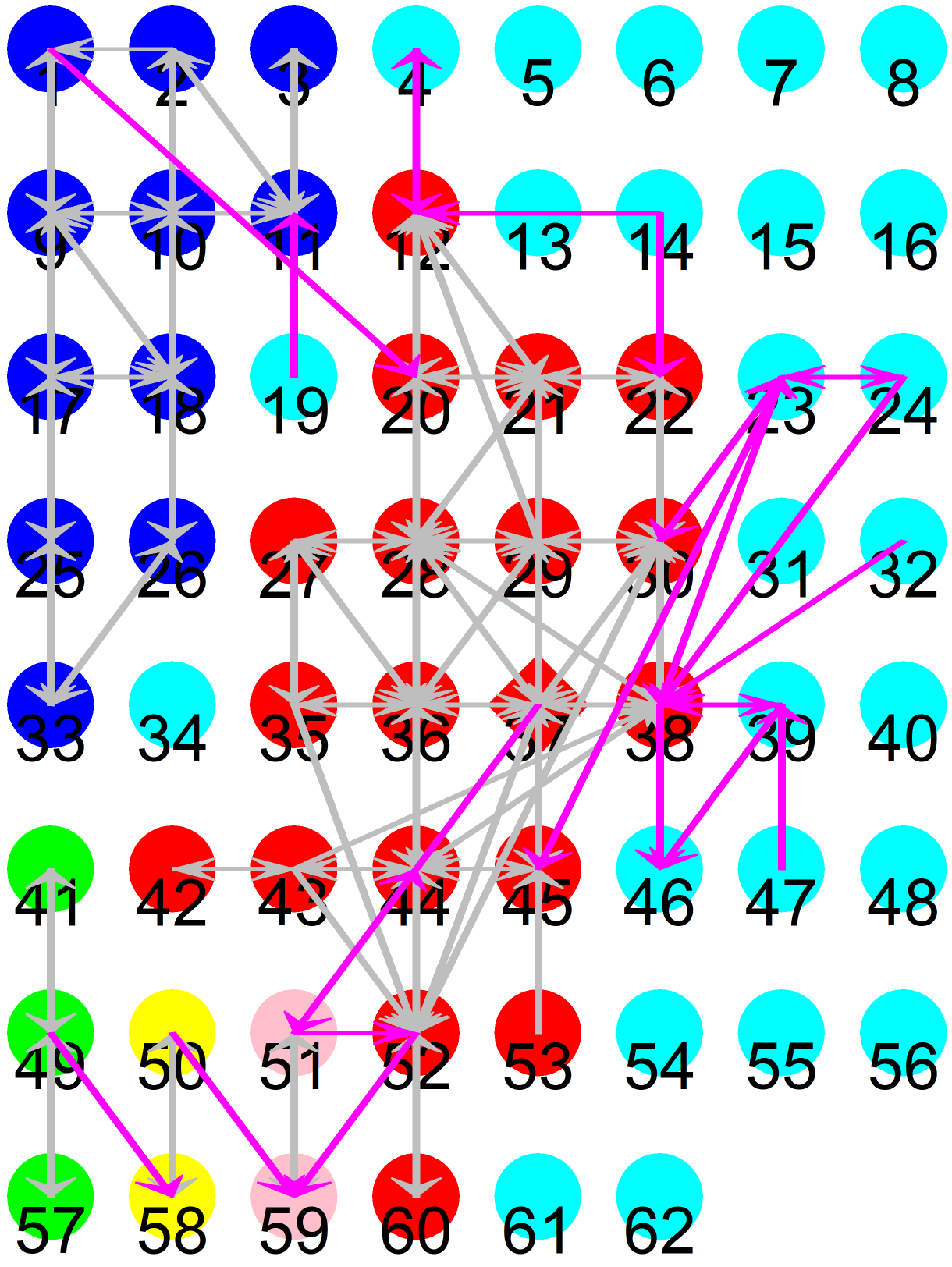}
}
 \subfigure[\footnotesize{$t\in [25 ,50]$ seconds}]
{
    \label{fig:after26_50}
    \includegraphics[width=0.23\textwidth,height=4.0cm,trim= 70mm 67mm 60mm 22mm,clip=TRUE]{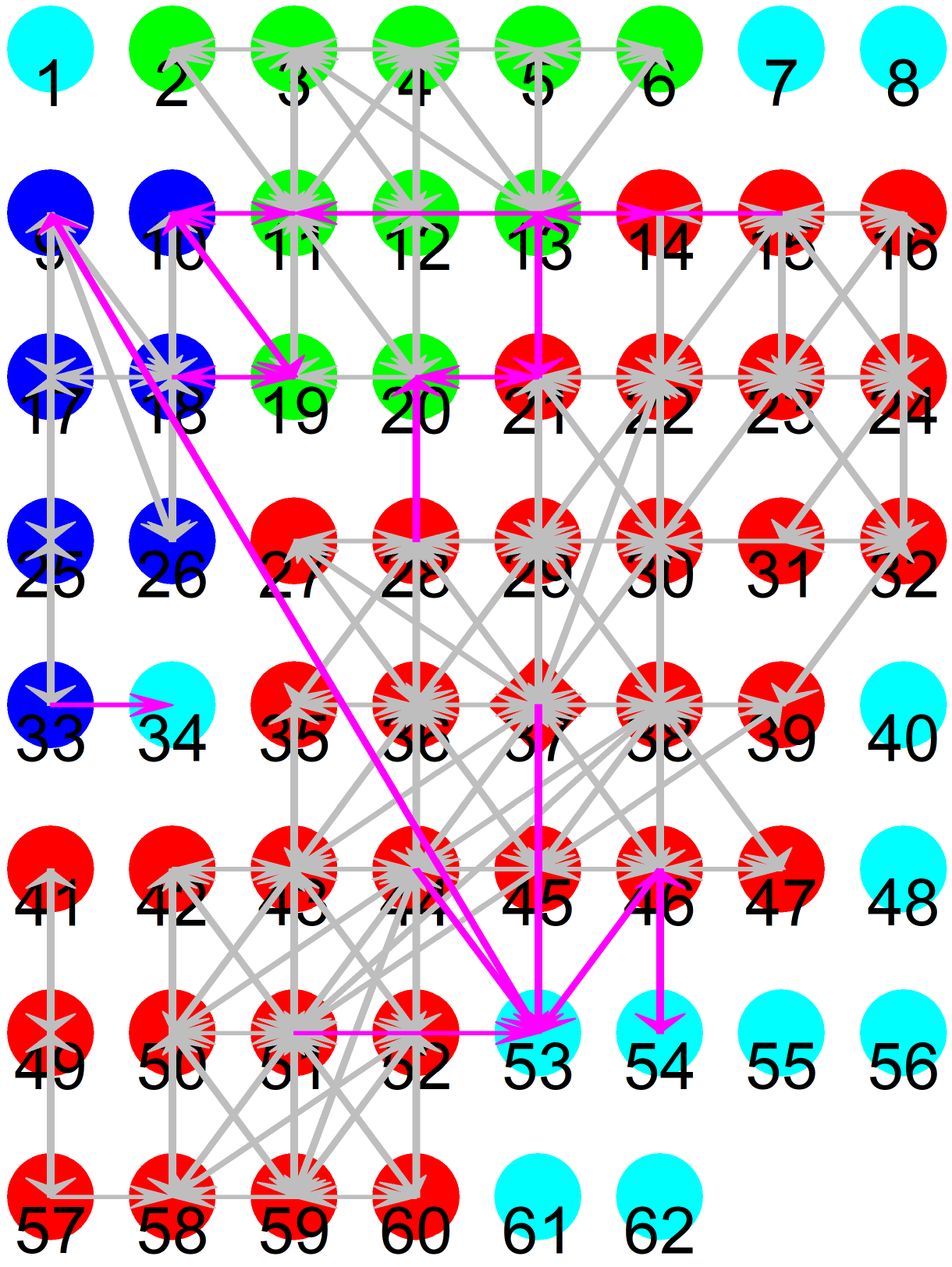}
}
\caption{{Brain networks for four periods}. \label{fig:BrainNetwork}\footnotesize{$t=0$ is the starting time of seizure onset. Grey and purple edges indicate within-cluster and between-cluster directional connections, respectively, based on a threshold corresponding to 1\% p-value. The node in the diamond is G37, the true SOZ. A node in light blue corresponds to a region in a cluster containing itself only. Nodes in the same color (dark blue, green, pink, red or yellow) are regions identified to be in the same cluster.}}
\end{figure}
\subsection{Network Results}
 Figures \ref{fig:before25_2}-\ref{fig:after26_50} show estimated networks for the four periods using the thresholds corresponding to the p-value of 1\%. The SOZ is at G37, indicated by the diamond in all these four figures, while all the other regions are indicated by circles. The shown network edges (in grey or purple) indicate their network edge probabilities above the threshold; and the nodes indicated by the same color other than light blue are corresponding to the regions identified to be in the same cluster. Each region indicated by light blue forms one cluster that contains the region itself only.

Our method reveals that the networks for the two pre-seizure periods were similar (Figures \ref{fig:before25_2} and \ref{fig:before25_1}), indicating that the subject's brain network was steady before seizure onset. However, dramatic changes occurred in the networks once seizure started (Figures \ref{fig:after1_25} and \ref{fig:after26_50}). Compared to the pre-seizure networks, more regions were connected to the SOZ (G37) and fell into the same cluster as the SOZ, indicating that the activity of the SOZ affected more and more regions as seizure developed. This result is in line with the existing understanding of seizure propagation \citep{englot2016regional,rosenow2001presurgical}.

\begin{figure}
\centering
\subfigure[\footnotesize{Corr. Pre-onset}] 
{    \label{fig:CorrBefore}
    \includegraphics[width=3.75cm,height=3.2cm,trim= 69mm 66mm 60mm  22mm,clip=TRUE]{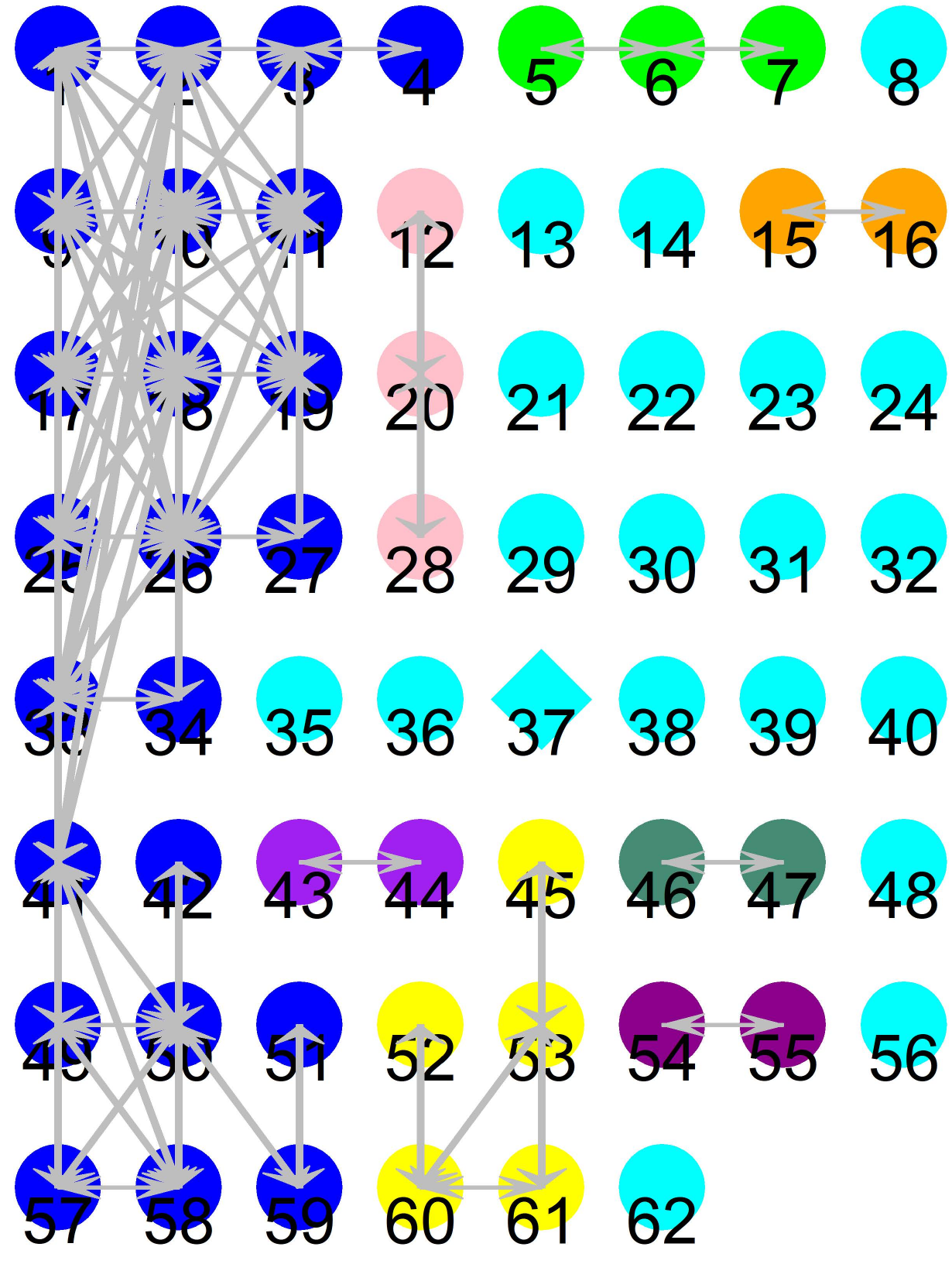}
}
\subfigure[\footnotesize{Corr. Onset}] 
{    \label{fig:CorrAfter}
    \includegraphics[width=3.75cm,height=3.2cm,trim= 69mm 66mm 60mm  22mm,clip=TRUE]{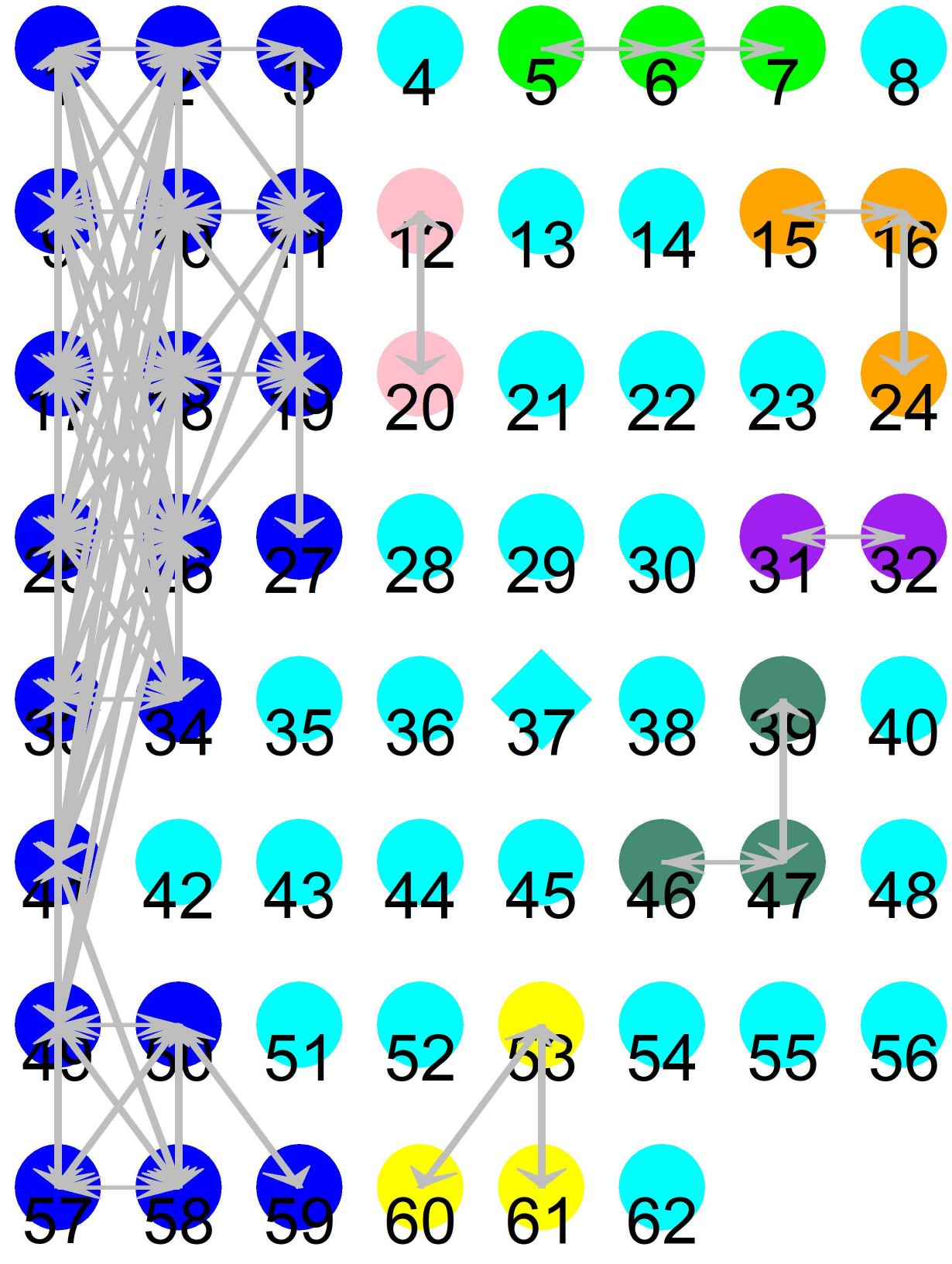}
}
 \subfigure[\footnotesize{Cross-Corr. Pre-onset}]
{
    \label{fig:pCorrBefore}
    \includegraphics[width=3.75cm,height=3.2cm,trim=69mm 66mm 60mm  22mm,clip=TRUE]{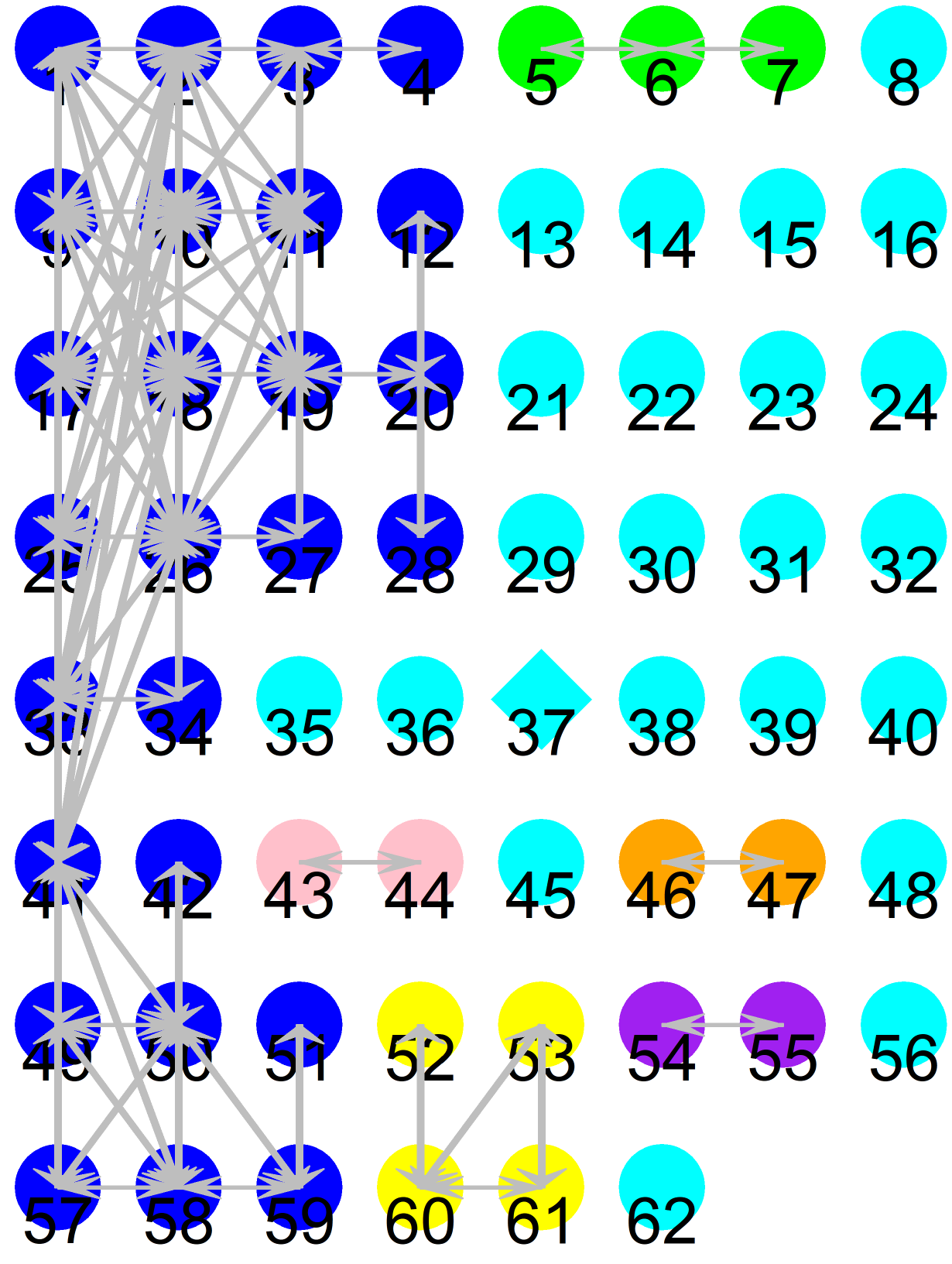}
}
 \subfigure[\footnotesize{Cross-Corr. Onset}]
{    \label{fig:pCorrAfter}
    \includegraphics[width=3.75cm,height=3.2cm,trim= 69mm 66mm 60mm 22mm,clip=TRUE]{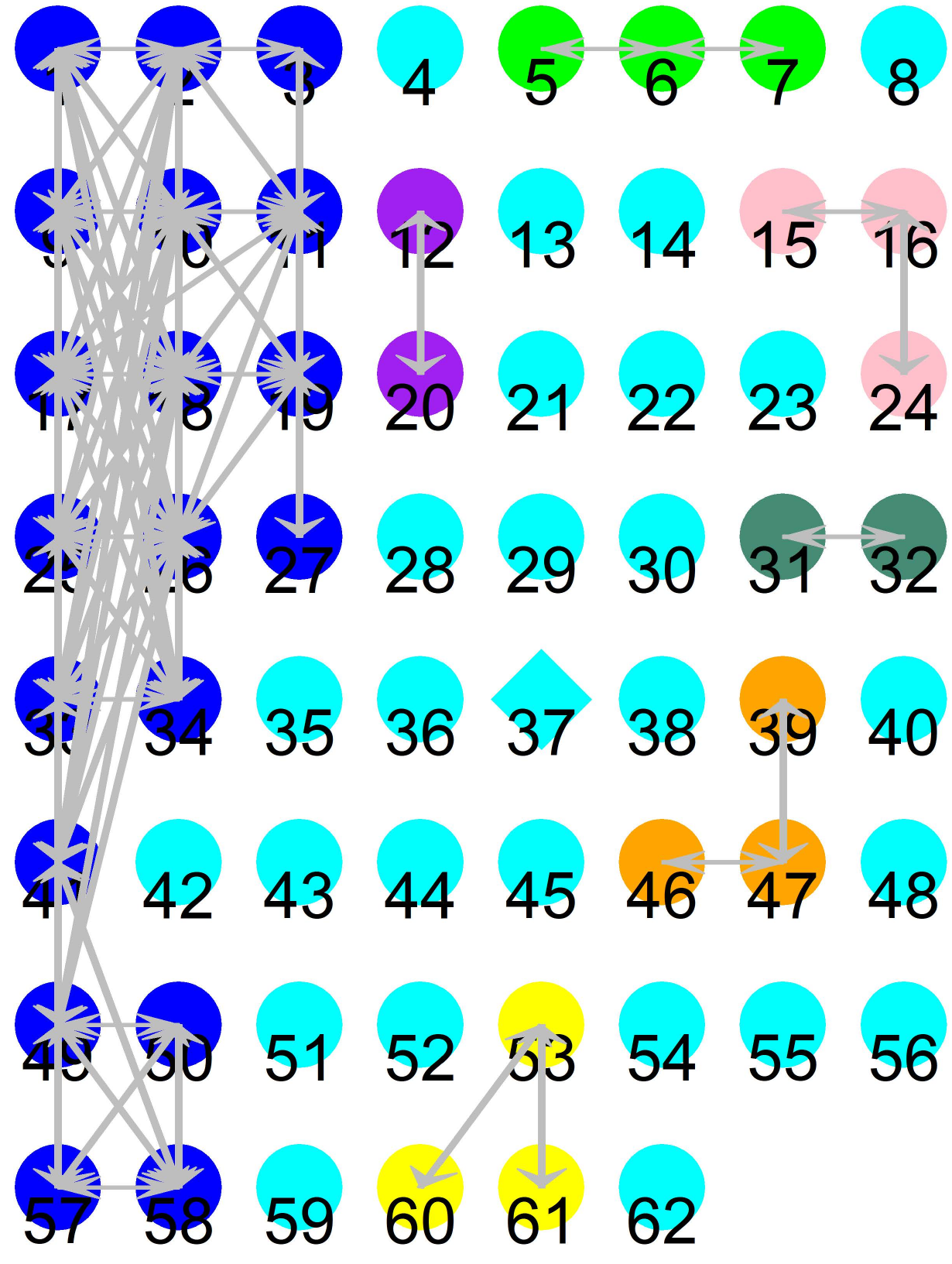}
}
\\
\subfigure[\footnotesize{MAR($L_1$) Pre-onset}] 
{    \label{fig:L1Before}
    \includegraphics[width=3.75cm,height=3.2cm,trim= 69mm 66mm 60mm 22mm,clip=TRUE]{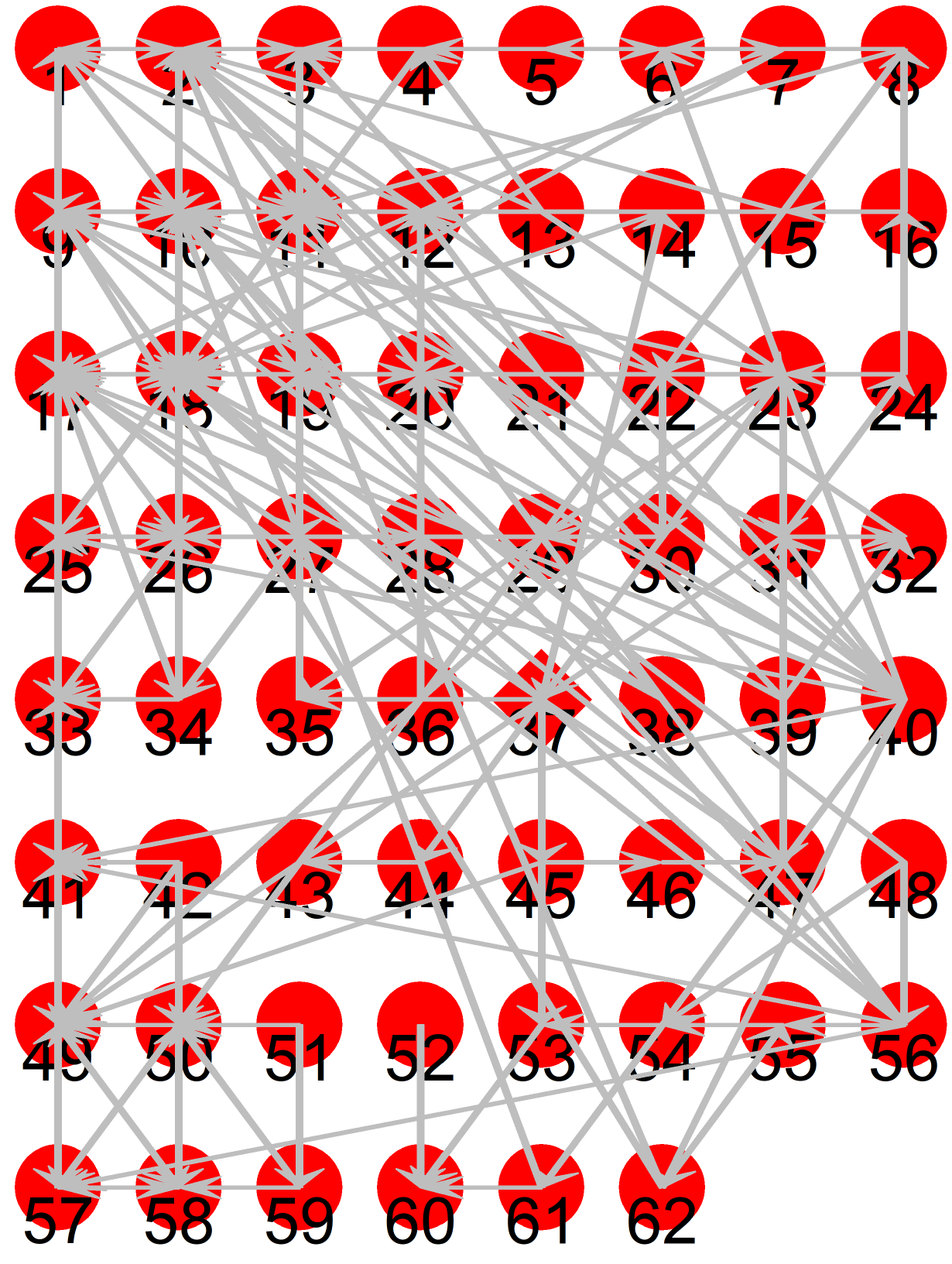}
}
\subfigure[\footnotesize{MAR($L_1$) Onset}] 
{    \label{fig:L1After}
    \includegraphics[width=3.75cm,height=3.2cm,trim= 69mm 66mm 60mm 22mm,clip=TRUE]{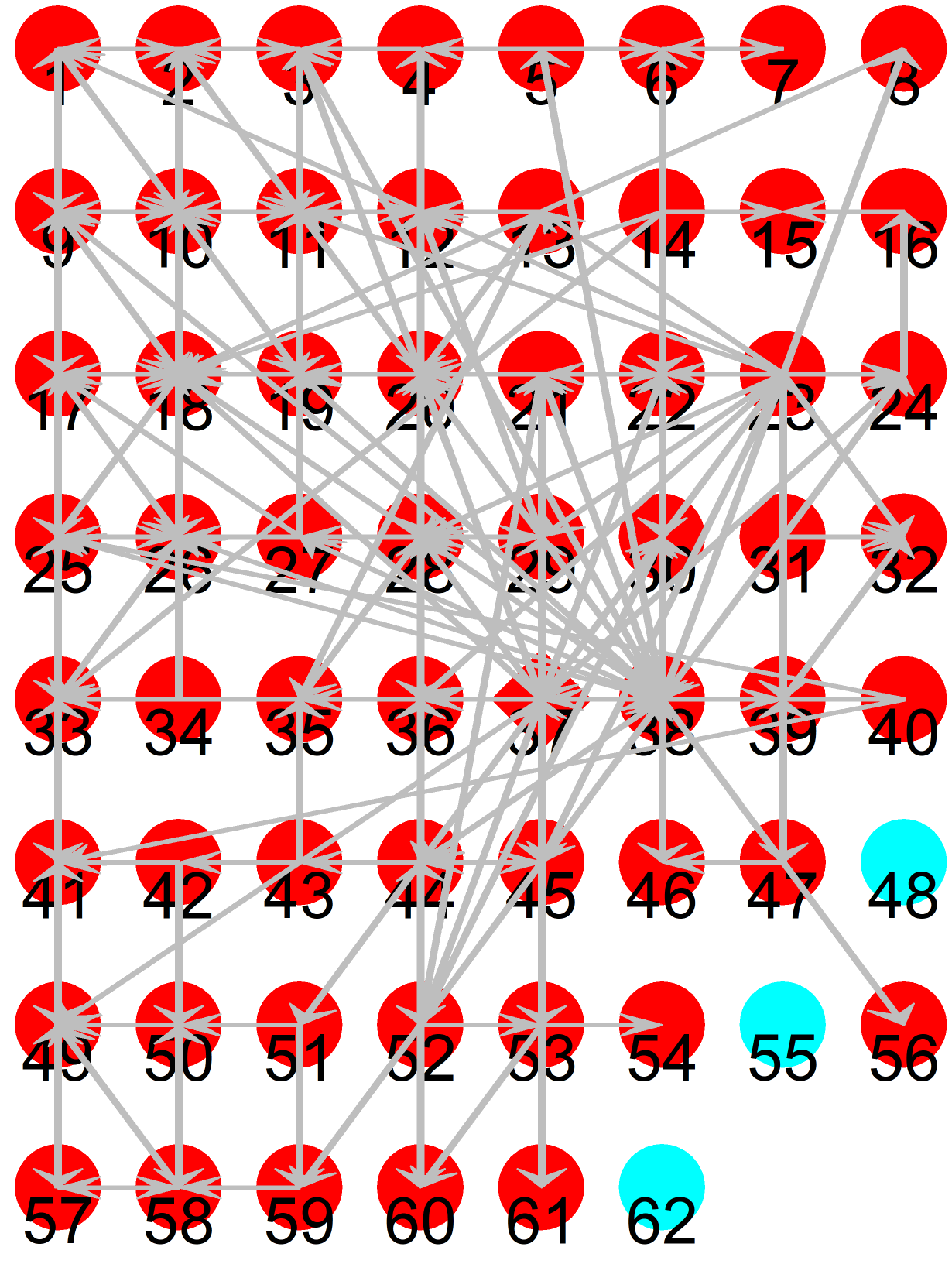}
}
 \subfigure[\footnotesize{Glasso Pre-onset}]
{
    \label{fig:GlassoBefore}
    \includegraphics[width=3.75cm,height=3.2cm,trim= 69mm 66mm 60mm 22mm,clip=TRUE]{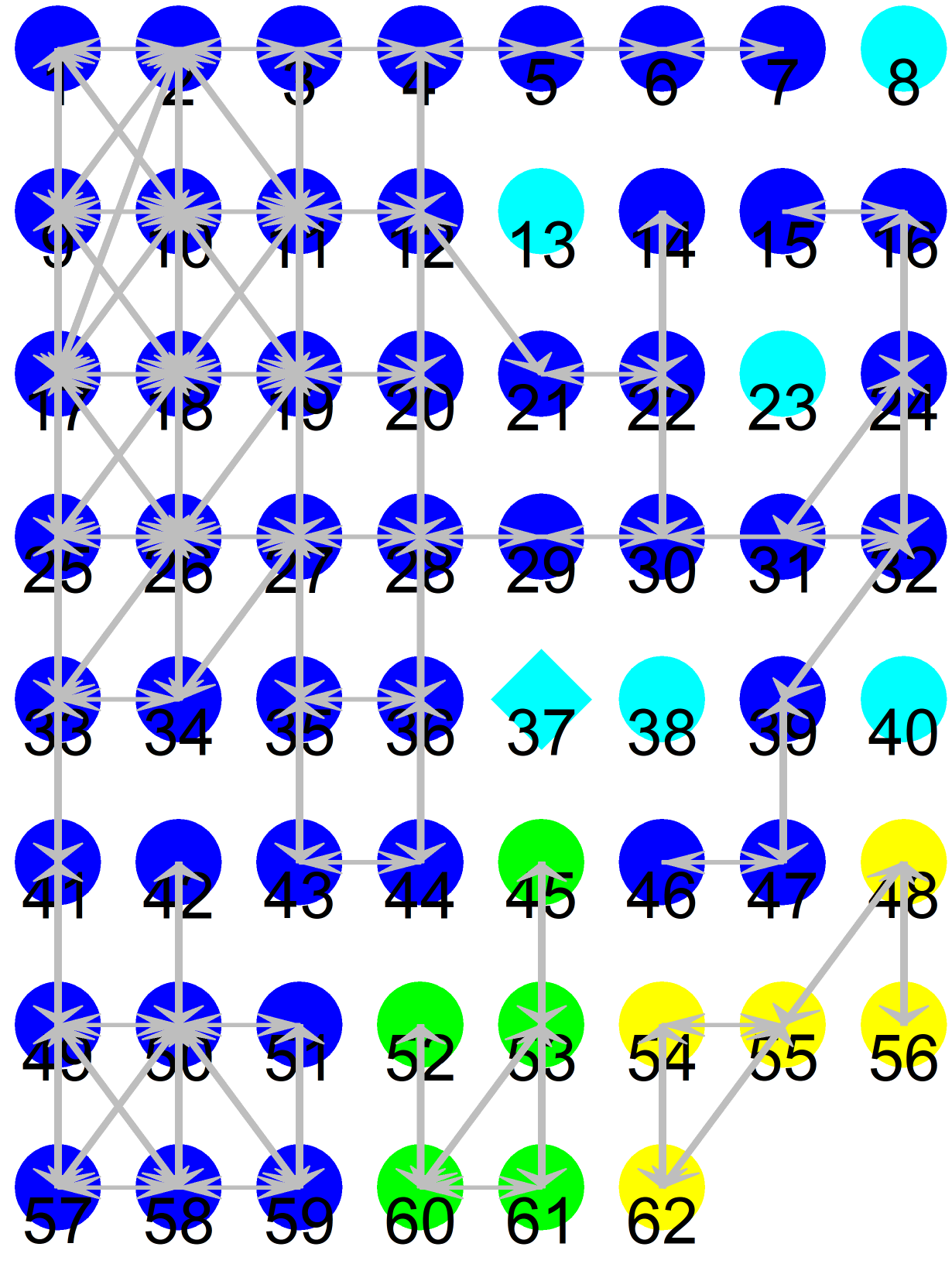}
}
 \subfigure[\footnotesize{Glasso Onset}]
{    \label{fig:GlassoAfter}
    \includegraphics[width=3.75cm,height=3.2cm,trim= 69mm 66mm 60mm 22mm,clip=TRUE]{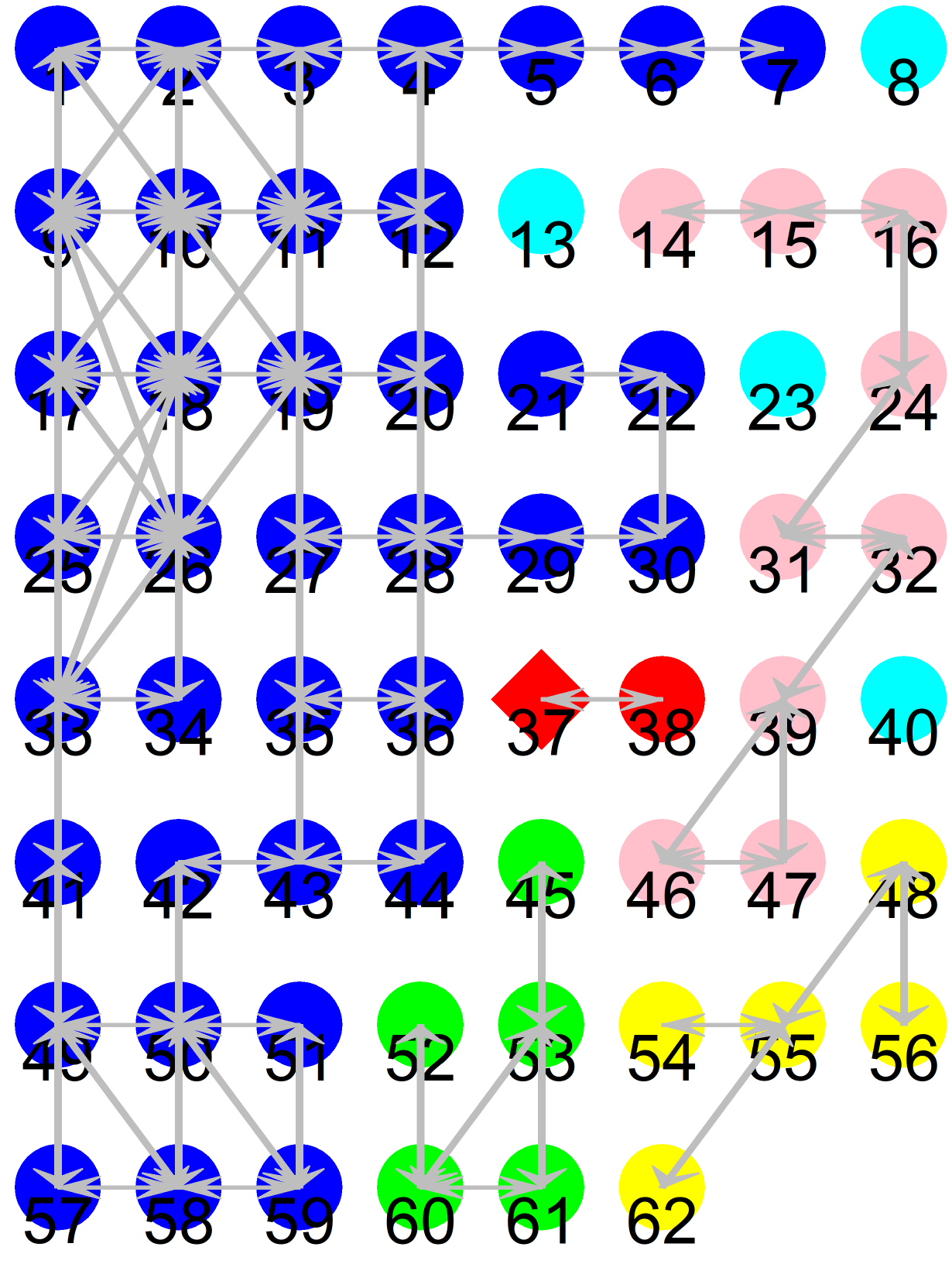}
}\\
\subfigure[\footnotesize{PDC Pre-onset}] 
{    \label{fig:PDCBefore}
    \includegraphics[width=3.75cm,height=3.2cm,trim= 69mm 66mm 60mm 22mm,clip=TRUE]{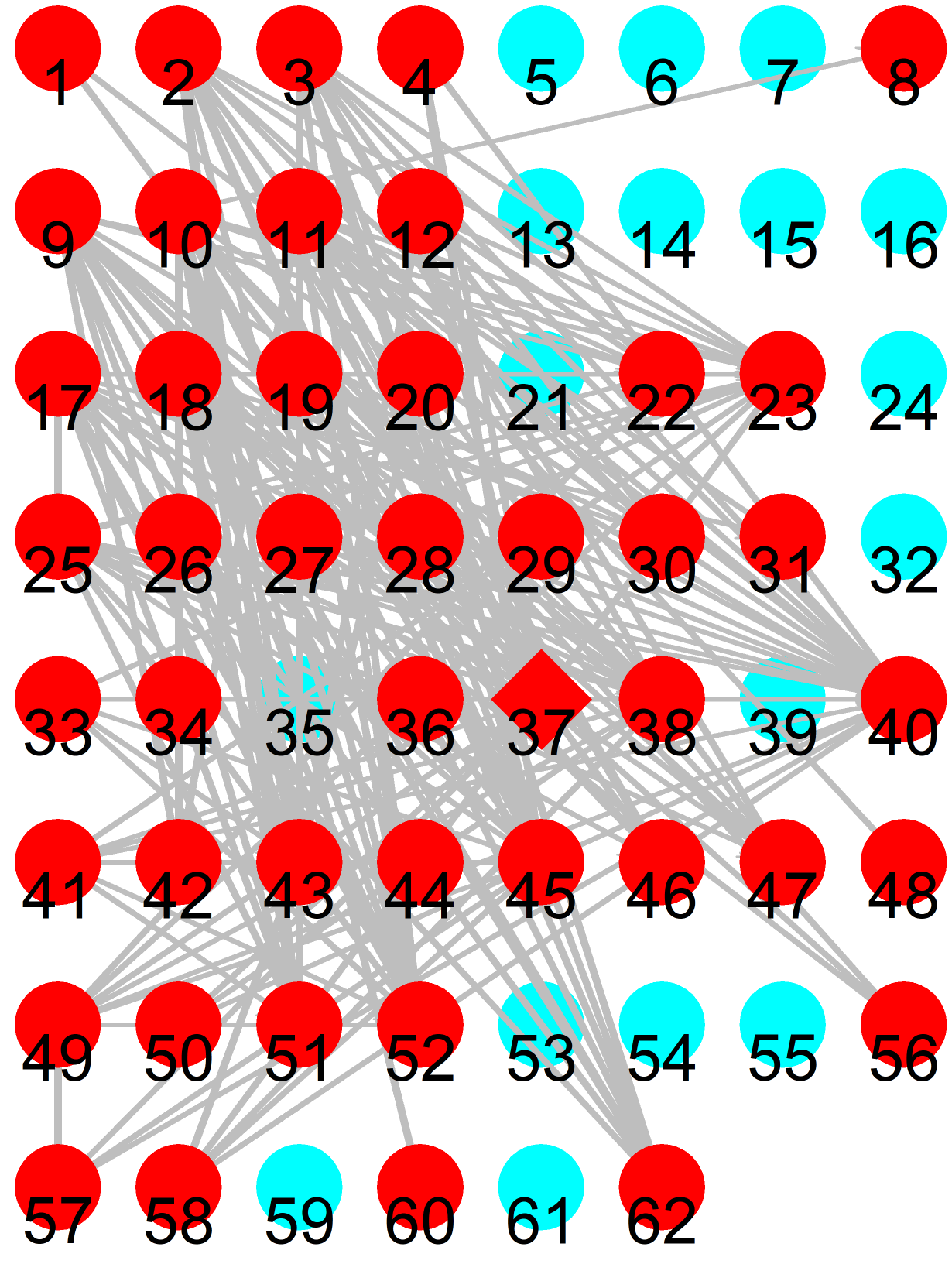}
}
\subfigure[\footnotesize{PDC Onset}] 
{    \label{fig:PDCAfter}
    \includegraphics[width=3.75cm,height=3.2cm,trim= 69mm 66mm 60mm 22mm,clip=TRUE]{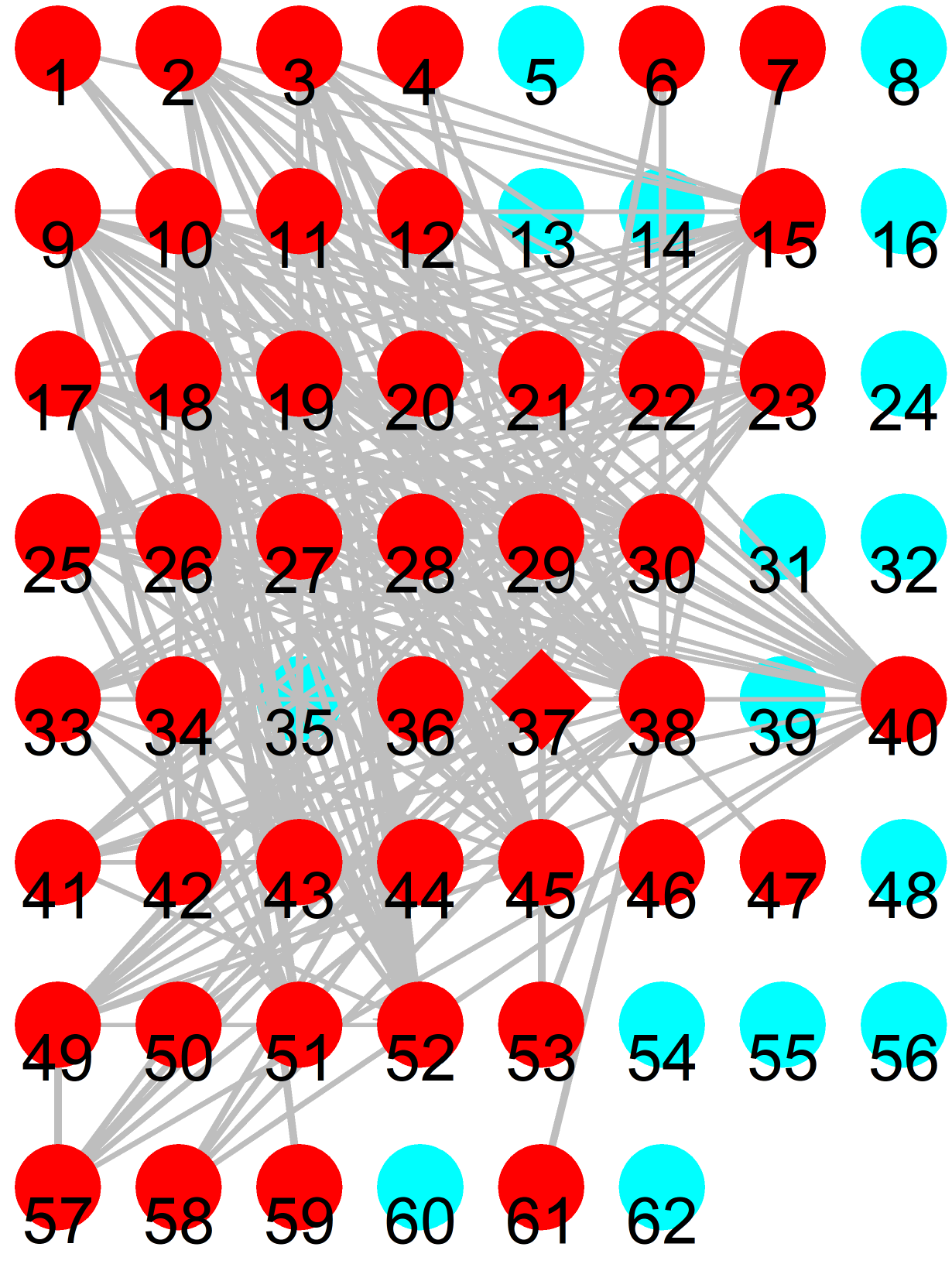}
}
 \subfigure[\footnotesize{DTF Pre-onset}]
{
    \label{fig:DTFBefore}
    \includegraphics[width=3.75cm,height=3.2cm,trim= 69mm 66mm 60mm 22mm,clip=TRUE]{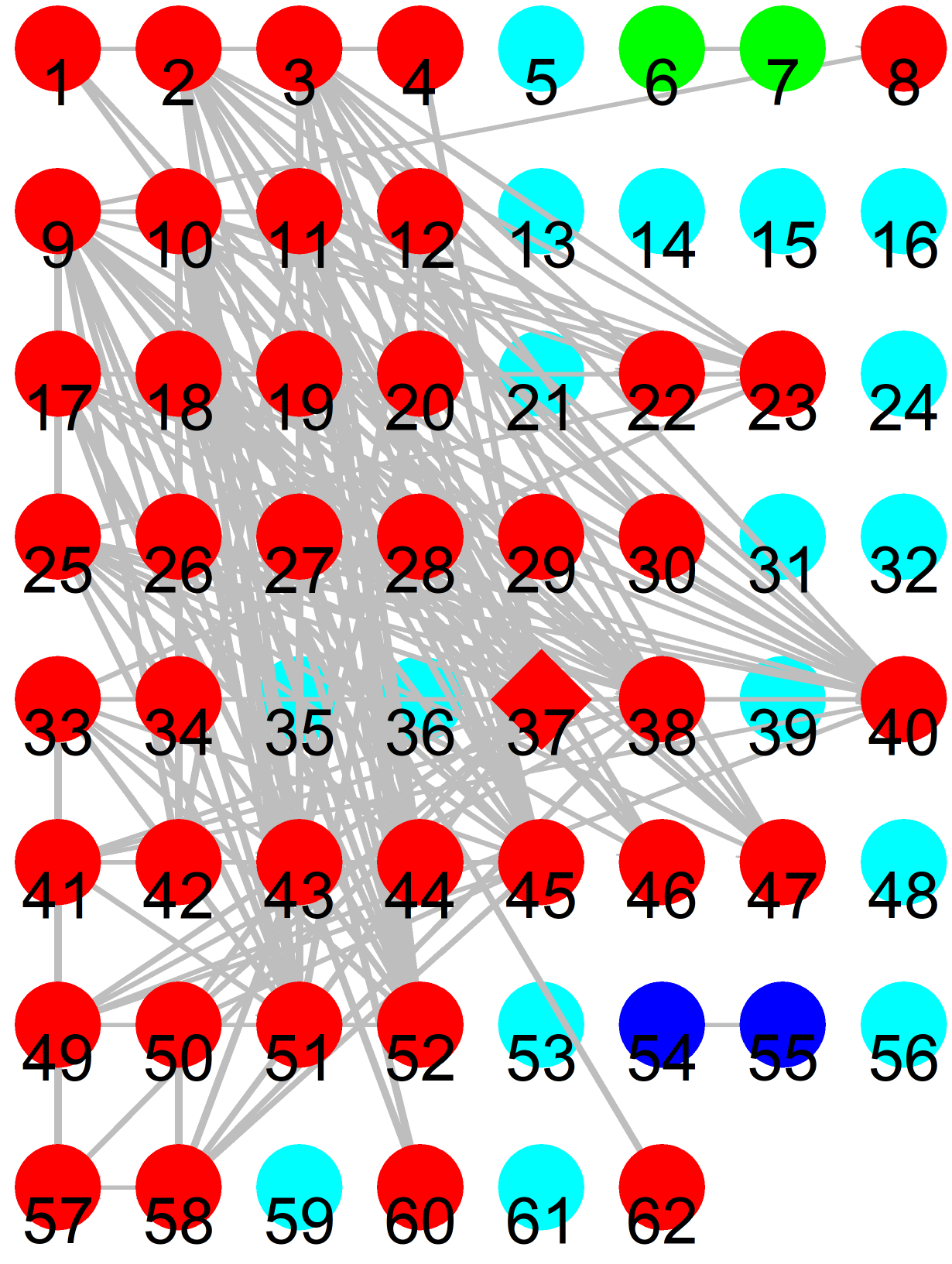}
}
 \subfigure[\footnotesize{DTF Onset}]
{    \label{fig:DTFAfter}
    \includegraphics[width=3.75cm,height=3.2cm,trim= 69mm 66mm 60mm 22mm,clip=TRUE]{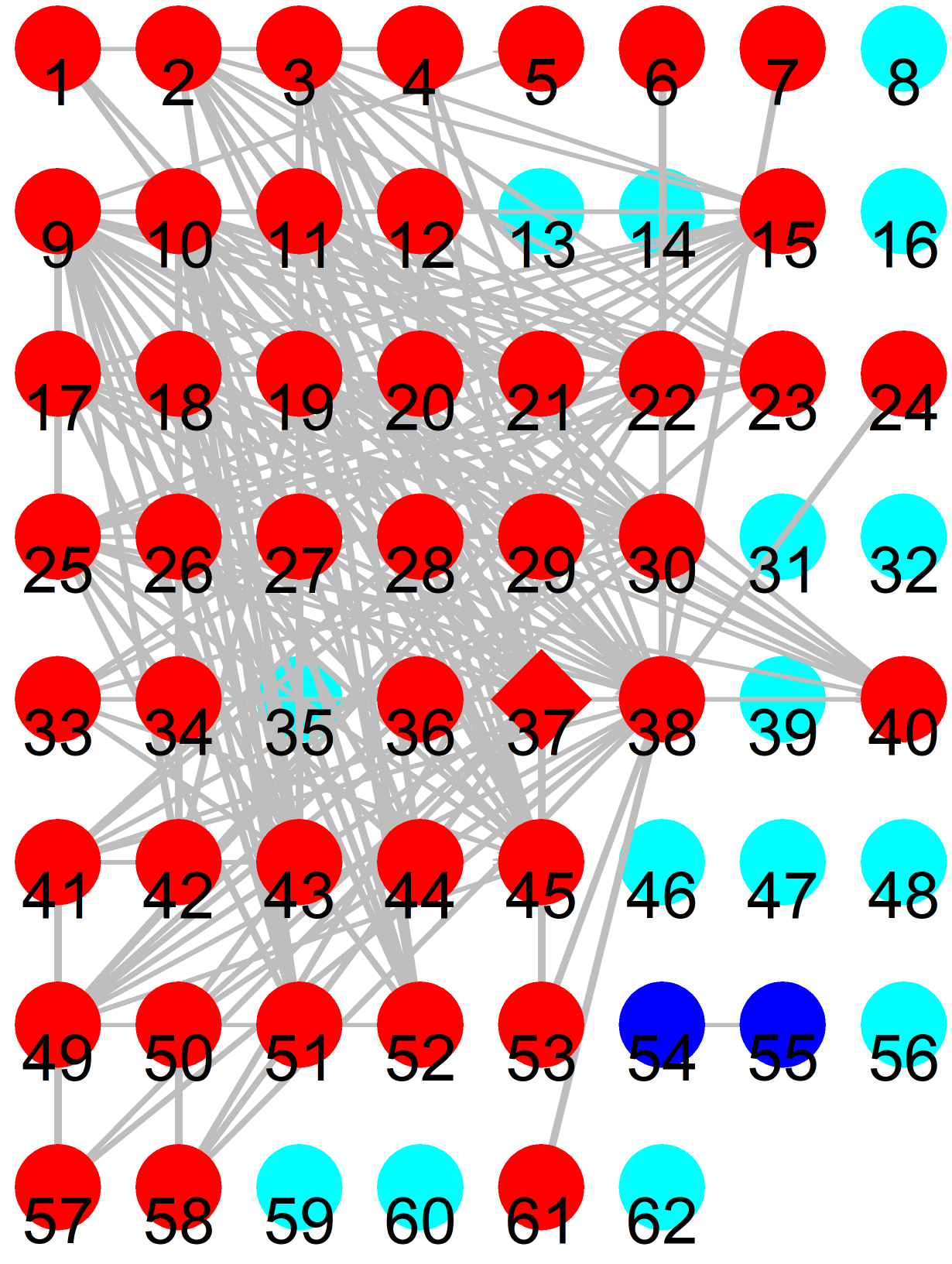}
}
\caption{\label{fig:BrainNetworkAlt}\small{Brain networks estimated using correlation, cross-correlation, MAR with an $L_1$ penalty (MAR($L_1$)), graphical Lasso (Glasso), partial directed coherence (PDC), and directed transfer function (DTF) methods.} {Each network edge indicates a pair of regions identified to be connected by the competing methods.} }
\end{figure}

To demonstrate the advantages of our method, we also analyzed the same iEEG data using several competing methods, including correlation, cross-correlation, partial directed coherence (PDC) \citep{baccala2001partial}, directed transfer function (DTF) \citep{Kaminski91}, $L_1$-penalized MAR (MAR($L_1$)), and graphical lasso (Glasso) \citep{friedman2014glasso,witten2011new}. We used each of these methods to analyze 300 1-second segments independently and obtained 300 calculated values for each candidate network edge (either directional or undirectional depending on the method). For each candidate network edge, we used the average of 75 values in each period to quantify the strength of connection. For comparison, we selected network edges with top 5\% averages, because the network edges selected by our method based on the p-value of 1\% roughly correspond to the edges with top 5\% $\bar{P}^\gamma_{ij}$s. Figures \ref{fig:CorrBefore}-\ref{fig:DTFAfter} show the networks estimated by the competing methods in the periods right before and right after the seizure onset. All these popular methods failed to detect the changes in the network at the seizure onset time, as evidenced by the similarity between the pre-onset and seizure-onset networks.

\subsection{SOZ Localization}
We hypothesize that the SOZ exhibits a significant change in its connectivity to other regions at the seizure onset. To quantify this change, we developed the following method. For each period, for each region, say $j$, we calculated the average of network edge probabilities from $j$ to all the other regions, $\sum_{i} \bar{P}^{\gamma}_{ij}/d$, referred to as region $j$'s average directional connectivity (ADC) in the period. We use the ADC difference between the periods right after and before the seizure onset to quantify the change in directional connectivity from region $j$ to other regions. Figure \ref{fig:ADC} shows the ADC changes of 62 regions at the seizure onset. Except for one region, the SOZ and its neighboring regions have the highest increases in ADC.

\begin{figure}[ht]
\centering
\subfigure[]
{    \label{fig:ADC}
    \includegraphics[width=9cm,height=4cm,trim= 0mm 1mm 0mm 0mm,clip=TRUE]{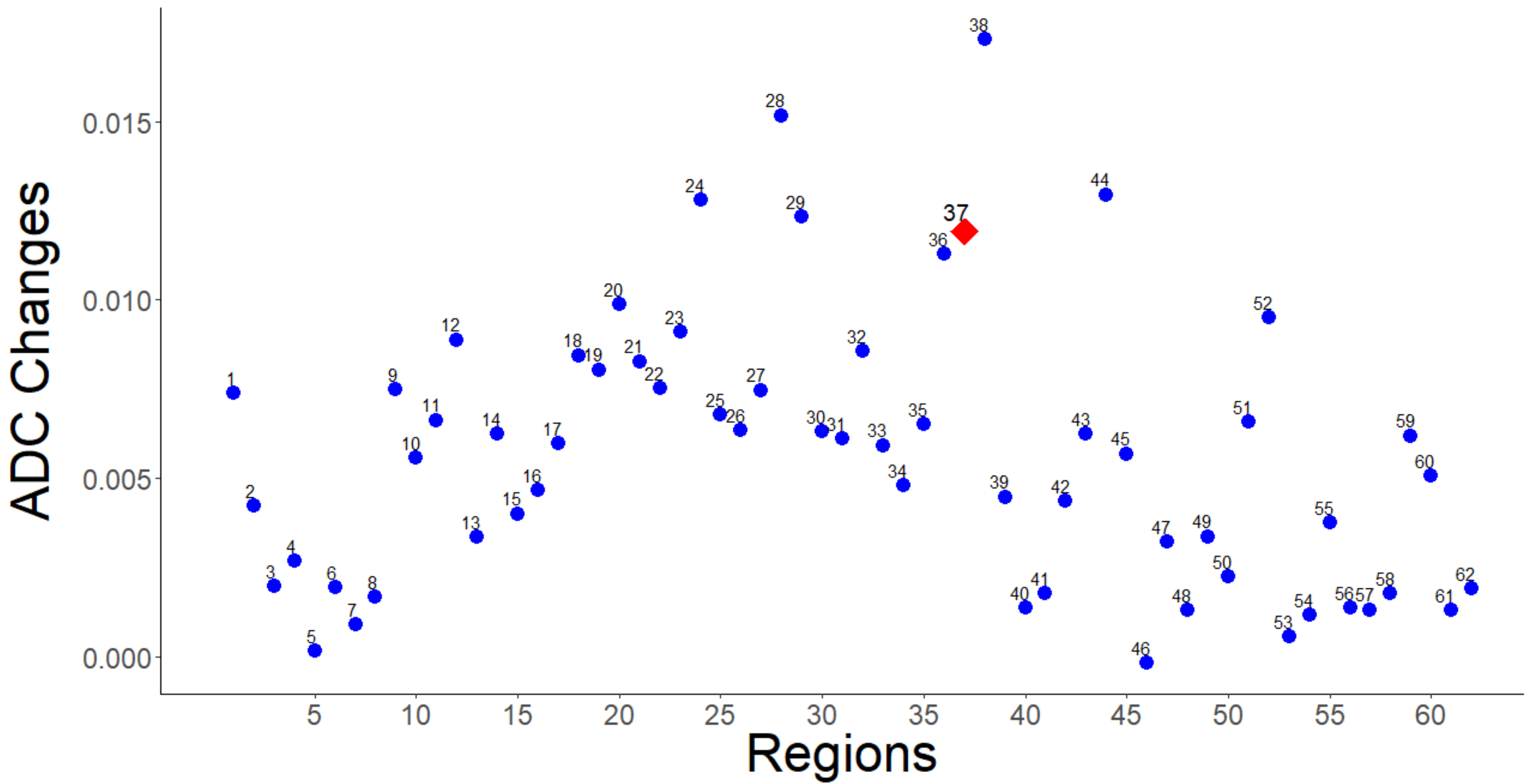}
}
\subfigure[]
{    \label{fig:SOZlocal}
    \includegraphics[width=0.25\textwidth,height=4cm,trim= 64mm 61mm 54mm 22mm,clip=TRUE]{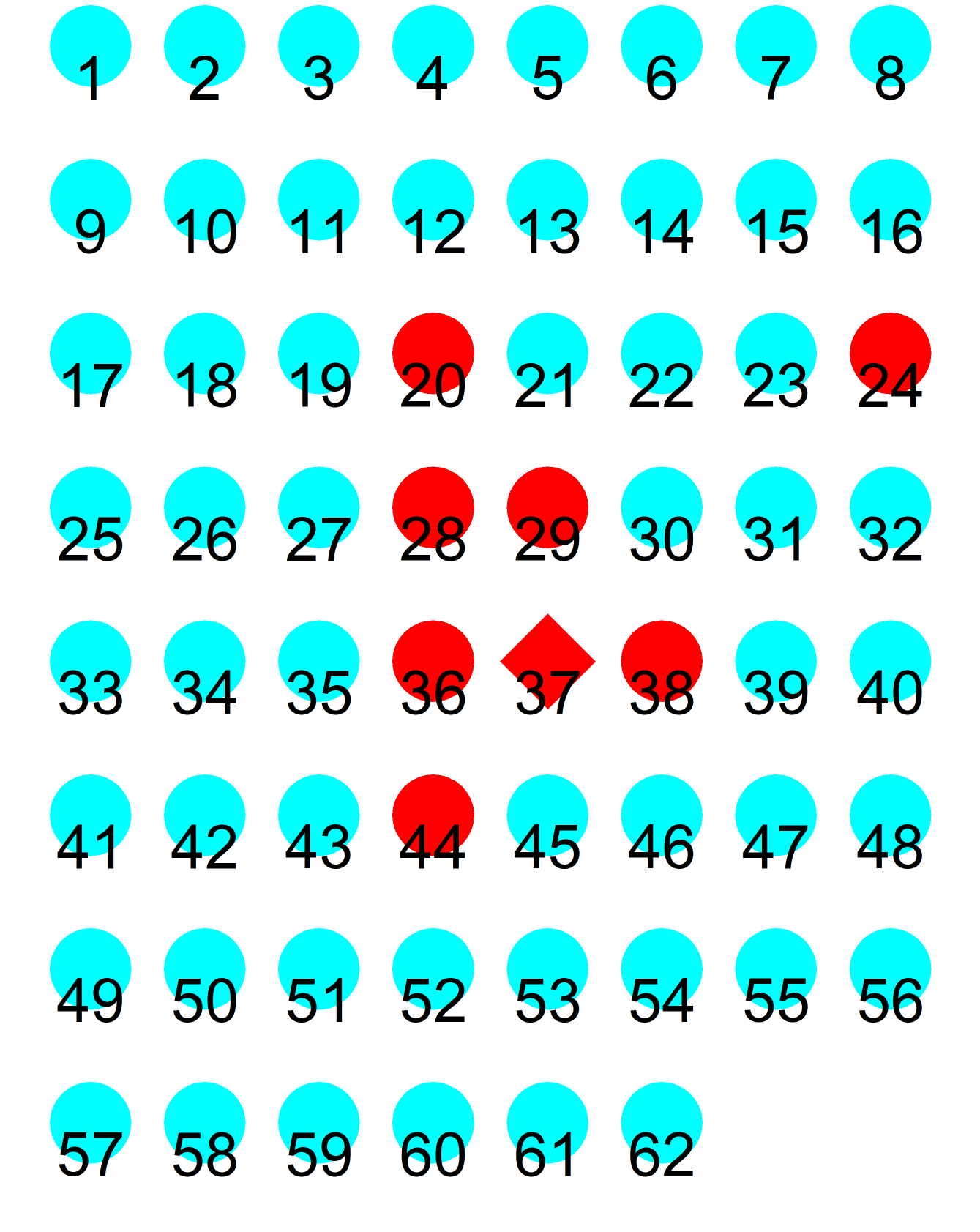}
}
\caption{\label{fig:ADC}{\footnotesize{ (a) Directional connectivity changes of 62 regions at seizure onset. (b) Regions with highest increases in directional connectivity. }}}
\end{figure}

We propose to select the regions with high ADC increases to be candidates for SOZ. To determine the threshold for ADCs, we calculated the 62 regions' ADC changes in the first two pre-seizure periods for the 3 seizures recorded by iEEG. Then we selected the regions whose ADC changes at the seizure onset are larger than the maximum of ADC changes in the two pre-seizure periods. Figure \ref{fig:SOZlocal} shows the selected regions (in red).

Our result showed that the small brain area including the SOZ G37 has the highest increase in directional connectivity at the seizure onset. This result is in line with the existing literature about the SOZ \citep{Engel94}: the abnormal, excessive neuronal activity starts from it and spreads to other regions. Our method quantified brain network changes and uncovered that the brain area including the SOZ first demonstrated an increase in directional connectivity during the seizure development. 

In summary, with our method, we revealed three characteristics of the epileptic patient's directional brain network. (1) The patient's network changed at the seizure onset time. (2) The change occurred around the SOZ, as the SOZ cluster expanded to include more regions, and the number of directional connections between the SOZ and other regions increased. (3) The extent of the directional connectivity of the SOZ increased most compared to other regions at the seizure onset time. These three results are in line with the existing understanding of seizure initiation and propagation. In contrast, existing network methods could not obtain the above three results together. These results are useful for identifying the brain areas affected by seizures and for evaluating the effect of seizures on brain functions. Also, our method has the potential to help clinicians localize the SOZ and, thus, to improve epilepsy diagnosis and treatment.


\section{Discussion}
This paper develops a new high-dimensional dynamic system model for mapping directional brain networks using iEEG data.
The proposed approach has three novelties. First, we propose a state-space first-order MAR-based model for the brain network. This model is effective for approximating various high-dimensional brain systems and is robust to violations of model assumptions. Second, in contrast to standard SSMAR and MAR models, the proposed Bayesian framework incorporates the prior knowledge of the cluster structure into the model estimation, which addresses the challenge in detecting connected brain regions among many possible ones. Our method produces scientifically meaningful network results. Third, we develop a stochastic-blockmodel (SBM)-motivated prior to impose the cluster structure on the SSMAR parameters that denote directional edges. This is novel from standard SBMs for observed networks where network edges are directly known.

The proposed method can robustly detect directional connections with high accuracy, even if the underlying model for the brain network is nonlinear for three reasons. First, we apply the SSMAR to short iEEG time segments so that the linear MAR can effectively approximate the underlying network system. Second, we use the proposed model to identify the directional connections through detecting the existence of temporal dependence among neuronal activities of regions rather than estimating the nonlinear interactions among regions. The first-order SSMAR focuses only on the primary temporal dependence (rather than the exact order or nonlinearity of the dependence) among multivariate time series. Thus, the model is parsimonious in terms of the number of model parameters and enables efficient detection of directional connections among many regions. Third, the SBM-motivated prior can effectively capture potential brain network patterns. Using the SBM-motivated prior increases the efficiency in detecting directional connections. In summary, the proposed integration of a conventional SSMAR and the cluster structure yields robustness, flexibility, efficiency, and computational feasibility in modeling and estimating brain network
systems.

We have applied statistical methods used for localizing the SOZ based on EEG data to our iEEG data. Specifically, \cite{schroder2019fresped} developed frequency specific methods to localize the SOZ through detecting changes in EEG data; and \cite{wang2018topological} used the differences in persistent homology between EEG data in pre-seizure and seizure periods to localize the SOZ. However, these methods tend to have much higher FPRs than the proposed method most likely because EEG and iEEG data have different properties. The two methods \citep{schroder2019fresped,wang2018topological} require the time series before and after seizures to be stationary for a relatively long period. Since the regions recorded by EEG are large and spatially distant from each other, the changes in one EEG region take a relatively long time to spread to other regions. As such, the assumption of stationary long time series required by the two SOZ localization methods can be satisfied with EEG data. In contrast, regions recorded by iEEG electrodes are spatially close. Seizures can propagate from the SOZ to other regions quickly, and thus, many regions surrounding the SOZ can have many sharp changes in frequencies and persistent homology in a short period of time. This phenomenon makes it difficult for the methods that rely on relatively long stationary time series to separate the SOZ from its many surrounding regions. Because our method is focused on detecting the change in directional connectivity instead of the change in time series, our method can better exclude non-SOZ regions between which directional connectivity remains unchanged at the seizure onset. 

\bibliography{SBSSMAR_Orig}

\end{document}